\pdfoutput=1
\documentclass[conference]{IEEEtran}
\usepackage[utf8]{inputenc}
\usepackage[T1]{fontenc}
\usepackage{graphicx}
\usepackage{booktabs}
\usepackage{tikz}
\usetikzlibrary{positioning,arrows.meta,shapes.geometric,fit}
\usepackage{xcolor}
\usepackage{url}
\usepackage[hidelinks]{hyperref}
\hypersetup{pdftitle={F(AI)2R: Who Did What, and Who Checked? Verifiable AI Provenance as an Executable Skill},
  pdfauthor={Florian Krebs}}
\newcommand{\MTriples}{3\,323}
\newcommand{\MActivities}{244}
\newcommand{\MClaims}{8}
\newcommand{\MSources}{75}
\newcommand{\MSrcChecked}{73}
\newcommand{\MSrcSelf}{5}
\newcommand{\MHumanRungs}{34}
\newcommand{\MCommits}{222}
\newcommand{\MBookCommits}{87}
\newcommand{\MTurns}{3\,284}
\newcommand{\MOutTok}{2\,551\,954}
\newcommand{\MInTok}{17\,098}
\newcommand{\MCacheReadM}{860.4}
\newcommand{\MCacheWriteM}{32.6}
\newcommand{\MCacheReadMint}{860}
\newcommand{\MCostUSD}{1641}
\newcommand{\MCostOpus}{820}
\newcommand{\MCostSonnet}{328}
\newcommand{\MOverReqPct}{12.6\%}
\newcommand{\MOverOutPct}{11.8\%}
\newcommand{\MOverCostPct}{9.0\%}
\newcommand{\MOverReqNum}{12.6}
\newcommand{\MOverOutNum}{11.8}
\newcommand{\MOverCostNum}{9.0}
\newcommand{\MBookCommitsPctNum}{39.2}
\newcommand{\MDirMsgs}{141}
\newcommand{\MDirWords}{2\,112}
\newcommand{\MDirMedian}{7}
\newcommand{\MActiveH}{12.7}
\newcommand{\MSpanH}{90}
\newcommand{\MPassAuth}{67}
\newcommand{\MPassAudit}{163}
\newcommand{\MPassRepair}{12}
\newcommand{\MPassBuild}{2}
\newcommand{\MProseWordsK}{9\,300}
\newcommand{\MToolLinesK}{2\,800}
\newcommand{\MSkillVersion}{v0.21}
\newcommand{\MGrantsIssue}{23}
\newcommand{\MGrantsSession}{13}
\newcommand{\MLinkOnly}{29}
\newcommand{\MPriceBasis}{input \$10, output \$50, cache read \$1, 1\,h cache write \$20}
\newif\ifextended
\extendedtrue

\begin{document}

\title{F(AI)\textsuperscript{2}R: Who Did What, and Who Checked?\\
Verifiable AI Provenance as an Executable Skill}

\author{\IEEEauthorblockN{Florian Krebs}
\IEEEauthorblockA{Deutsches Zentrum f\"ur Luft- und Raumfahrt (DLR)\\
ORCID: 0000-0001-6033-801X}}

\maketitle

\begin{abstract}
F(AI)\textsuperscript{2}R is FAIR research with AI in the loop,
twice: an AI-assisted \emph{authoring} pass and a machine-readable
\emph{audit} pass over every artefact. AI systems now draft, refactor,
and verify research artefacts, yet their
contributions are rarely recorded in a form a later human or machine can
audit. Building on the original F(AI)\textsuperscript{2}R
experiment, we generalize its provenance model beyond scholarly writing into
\emph{aiprov}, a PROV-O extension covering any AI-in-the-loop artefact,
and we package the method as an \emph{executable skill} that an AI agent
operates itself: setup asks the human operator for their ORCID iD,
resolves their identity from the public registry, and scaffolds continuous
integration that gates every push on graph conformance and publishes the
current build of this very paper. The paper is its own case study. Every
activity, claim, and source in its production is recorded in the
repository's provenance graph under two invariants: no parentless claim,
and verification rungs that only humans may grant.
\end{abstract}

\begin{IEEEkeywords}
AI provenance, PROV-O, research integrity, verification, agent
skills, FAIR, human oversight
\end{IEEEkeywords}

\section{Introduction}\label{sec:intro}

F(AI)\textsuperscript{2}R is not an established term, and that is
deliberate. It reads as FAIR~\cite{fair2016} with the AI factor
squared, because AI enters research twice: once \emph{authoring} an
artefact, and once \emph{auditing} it into a machine-readable record
of who did what, when, and from which sources. This paper defines that
double role precisely, generalizes it beyond scholarly writing, and
demonstrates it on its own production.

AI agents now draft manuscripts, refactor code, sweep literature, and
verify results. The debate about whether they should is loud and
well documented~\cite{chatgpt-wrote-it2023, chatgpt-priorities2023};
the record-keeping has not kept pace. A git commit stores \emph{what}
changed and a free-text author name; it does not record whether that
author was a human or a model, at what cost the change was made, under
which instructions, or from which sources. The work is real, but its
context evaporates at commit boundaries. Existing responses are either
policy (disclosure statements, authorship bans) or documentation
artefacts written after the fact, such as model
cards~\cite{modelcards2019}; neither is machine-checkable at the
granularity of a single claim.

We treat this as a question of scientific ethics, and we state our
position plainly. When AI co-writes, the ethical question is not
``is that allowed?'' It is: where does this come from, and can anyone
check it? Integrity then becomes a property of the record, not of the
author's assurances. Our stance is a measured one: not banning AI from
research, and not trusting it blindly, but building the layer on which
trust can be grounded. Provenance is that layer. A verification ladder
says who may vouch for what, and its top rungs are reserved for
humans. The incentive system makes this urgent rather than optional:
publish-or-perish pays for counted output, and hypercompetition
selects for whatever maximizes the
count~\cite{perverse-incentives2017}. Generative AI collapses the
marginal cost of exactly the thing the metric counts, onto a review
system already strained by output
growth~\cite{publishing-strain2024}. Section~\ref{sec:dilution} describes the predictable result. Appeals
to ethics alone do not bend that curve. Infrastructure that makes
verifiable work cheaper than unverifiable work can. The most expensive
mistake would be to scale speed without traceability.

This is a fast paper standing on slow work. The ideas were sharpened
across earlier experiments: \emph{Obscurity-Is-Dead}
\cite{obscurity-is-dead} established transcript-as-artifact and
per-claim verification labels while measuring how large language
models compress reverse-engineering effort, and the original
F(AI)\textsuperscript{2}R experiment~\cite{fair2r-repo} formalized the
two-pass model and the no-parentless-claim rule on a single scholarly
paper, bound to paper writing. Several strands of that programme
advance in parallel under one operator: this method, this paper, and a
system-integration fork of the shepard data-management
platform~\cite{shepard-fork}. That is precisely why per-activity
accounting matters: parallelism without provenance is how
contributions and errors alike become untraceable.

The paper makes three contributions. First, \emph{aiprov}, a
domain-agnostic generalization of the F(AI)\textsuperscript{2}R
vocabulary over PROV-O~\cite{prov-o2013}: agents, passes, claims, full
inference telemetry, and a verification ladder with human-only rungs
(Section~\ref{sec:method}). Second, the method packaged as an
\emph{executable skill} that an AI agent operates itself, from an
ORCID-first setup through continuous integration that gates every push
on graph conformance and publishes the current build of this very
paper (Section~\ref{sec:skill}). Third, a meta-experiment: the paper
is its own case study, and its provenance graph, including mishaps and
mid-experiment method changes, is the evaluation object
(Section~\ref{sec:meta}). Sections~\ref{sec:background}
and~\ref{sec:dilution} position the work and motivate it;
Sections~\ref{sec:discussion} and~\ref{sec:conclusion} discuss limits
and close the loop.

\section{Background and Related Work}\label{sec:background}

\textbf{Provenance models.} W3C PROV-O~\cite{prov-o2013} is the
standard vocabulary for expressing who generated what, through which
activity; it consolidated the lineage begun by the Open Provenance
Model~\cite{opm2010}, and PAV refined it for authoring and
versioning~\cite{pav2013}. Carina Haupt's overview of provenance use
cases documents the practice inside DLR, including monitoring software
development processes with provenance
data~\cite{provenance-overview2022}. aiprov extends this stack rather
than forking it: every class subclasses a \texttt{prov:} term, so
generic PROV tooling keeps working.

\textbf{Contributor attribution.} CRediT names who contributed what
kind of work~\cite{credit2019} and is now a NISO
standard~\cite{credit-niso2022}, but it operates at manuscript
granularity and admits only humans. aiprov records attribution
per activity and per claim, and its agent model includes AI systems
and deterministic tools alongside people.

\textbf{AI transparency artefacts.} Model cards~\cite{modelcards2019}
and datasheets~\cite{datasheets2021} document the \emph{model} or the
\emph{dataset}. What they do not capture is the collaboration that
produced a specific artefact: which prompts, which sessions, which
verification steps. RO-Crate~\cite{rocrate2022} is complementary
packaging; an aiprov graph can travel inside a crate. Closest to this
work, PROV-AGENT extends W3C PROV to capture agent interactions,
prompts, responses, and decisions, in agentic workflows in near real
time~\cite{prov-agent2025}; it shares the substrate and the
agent-centric telemetry, and differs in what this paper adds on top:
verification rungs reserved for humans, packaging as an executable
skill, and the method applied to its own production.
\ifextended
The differentiation deserves precision, since the substrate is
shared. PROV-AGENT instruments \emph{capture}: it wires agent
frameworks so that prompts, responses, and tool decisions land in a
PROV graph as they happen, serving debugging and reliability analysis
of agentic pipelines. aiprov adds the \emph{governance} layer that
capture alone does not provide: two enforced invariants, no
parentless claim and human-only rungs that no AI agent may grant;
claims as first-class resources whose verification state is part of
the record rather than a property of a log line; and a conformance
validator that fails a build instead of merely describing it. Capture
answers what happened; governance answers what may enter the record
and who is allowed to vouch for it. The two compose: a PROV-AGENT
style capture layer could feed an aiprov-governed record without
changing either vocabulary.

\textbf{FAIR digital objects and metadata infrastructures.} The FAIR
Digital Object line wraps research artefacts into typed, identifier-bearing
units with machine-actionable metadata~\cite{fairdo2020}; the
Helmholtz Metadata Collaboration standardizes the kernel attributes
such objects carry in its Kernel Information
Profile~\cite{hmc-kip2022}; and the Helmholtz Knowledge Graph
harvests exactly such metadata across siloed
infrastructures~\cite{unhide-helmholtz-kg}. An aiprov record is
FAIR-DO-shaped by construction, typed entities under persistent
identification (ORCID for people, DOI for sources, content hashes
and commits for artefacts), which is what makes the
infrastructure-level integration sketched in
Section~\ref{sec:conclusion} a binding exercise rather than a
redesign.
\fi

\textbf{The trust gap.} The large language models that now write
research prose descend from the transformer
architecture~\cite{attention2017}, and they fabricate plausible
citations~\cite{ai-hallucination2023}: existence of a reference is
machine-checkable, support for a claim is a judgement, and the two
must not be conflated. The ladder of Section~\ref{sec:method} keeps
them apart and names the grantor of every judgement.

\textbf{Neighboring lines beyond provenance vocabularies.} Claims
as first-class citizens predate this work: nanopublications make a
single assertion with its provenance a citable
object~\cite{nanopub2010}, and micropublications model claims,
evidence, and arguments formally~\cite{micropub2014}. aiprov
inherits that stance and adds what the AI era demands: the agent
class behind every activity, and a verification ladder whose grantor
is explicit and whose top rungs are human-only. On the artefact
side, C2PA Content Credentials sign machine-readable provenance
manifests for media, including AI-generated
content~\cite{c2pa-spec}, and in-toto binds software artefacts to
their supply-chain steps with signed
attestations~\cite{in-toto2019}. Both certify what a pipeline did
to bytes; aiprov records how humans and AI divided the research work
and who vouched for which claim, the layer the dual-witness ask of
Section~\ref{sec:discussion} would connect to this signing
infrastructure. Manubot demonstrated CI-built manuscripts with
citation-by-identifier years earlier~\cite{manubot2019}. These four
lines entered the record through the review that requested them:
the original novelty sweep, two registry queries, missed all of
them, the bounded-novelty mechanism working as designed, a narrow
recorded horizon visibly widened.

\textbf{Lineage and neighbors.} This work has three roots. The first
is AI-assisted research practice: \emph{Obscurity-Is-Dead}
\cite{obscurity-is-dead} introduced transcript-as-artifact and the
verification labels (\texttt{repo-vendored}, \texttt{lit-read},
\texttt{unverified-external}) that the present ladder canonicalizes,
and F(AI)\textsuperscript{2}R~\cite{fair2r-repo} formalized the
two-pass model and no-parentless-claim. The second is engineering
provenance: adjacent DLR work extends PROV-O with uncertainty
quantification for traceability in engineering
systems~\cite{uncertainty-prov2026}; the axes are complementary, since
aiprov records who and how, while uncertainty-aware provenance records
how confident, and a combined profile is future work
(Section~\ref{sec:conclusion}). The argument shape itself is
established in research software engineering: better architecture
makes better software makes better
research~\cite{better-architecture2025}; aiprov extends that chain
from the software to the record of its making. The third root is industrial semantic
modeling: the author co-authored the Asset Administration Shell
(AAS) Part~1 specification~\cite{aas-part1-v30rc02}, continued as
IDTA-01001~\cite{aas-part1-idta}. Where partners exchange information
across a value chain, semantics must be formal and machine-readable or
interoperability fails. aiprov applies that discipline to a new
exchange relationship: the one between humans, AI agents, and the
future auditors of their joint work. The Helmholtz Metadata
Collaboration~\cite{hmc, hmc-conference2025} provides the community
context: a provenance graph is metadata about \emph{work}, extending
FAIR metadata practice from data description to collaboration
description.

\section{The Dilution Crisis}\label{sec:dilution}

Scientific output has been growing faster than the community's
capacity to review it~\cite{publishing-strain2024}, and paper mills
industrialized fabrication well before language
models~\cite{paper-mills2021}. What changed is the cost curve.
\emph{Obscurity-Is-Dead} measured how language models compress the
effort gap for reverse-engineering proprietary
devices~\cite{obscurity-is-dead}; the same compression applies to
producing a plausible paper. GPT-fabricated papers are already
indexed and ranked by scholarly search
engines~\cite{gpt-fabricated2024}.

The demand side is the incentive system. Publish-or-perish pays for
counted output, and hypercompetition selects for whatever maximizes
the count~\cite{perverse-incentives2017}. Language models did not
create that pressure; they gave it a zero-cost supply channel.
Dilution is misaligned incentives meeting free volume.

The consequence is that finding good sources is now the bottleneck of
research. Retrieval trusts proxies: indexed, cited, fluent. Generation
has learned to imitate all three. Language models fabricate
plausible references and make citation errors at documented
rates~\cite{fabricated-citations2023, ai-hallucination2023}; once such
a reference is indexed, nothing in today's retrieval chain stops it
from being cited onward, and models trained on the polluted pool
degrade further~\cite{model-collapse2024}, so dilution compounds. The
parallelism that lets one operator advance several strands equally
industrializes pollution (Section~\ref{sec:discussion});
per-activity accounting is what tells the two apart.
Research infrastructures already treat
metadata quality as their central lever~\cite{hmc-conference2025},
and the deployed remedy is graph-shaped: the Helmholtz Knowledge
Graph harvests metadata from siloed infrastructures precisely to
drive better metadata practice~\cite{unhide-helmholtz-kg}. The
diluted literature is, in our words, a data swamp one level up, and
it admits the same remedy.

When text is cheap, the scarce good is an original idea whose support
is checkable. We treat originality as a provenance property: a novelty
claim carries its recorded search horizon, that is, what was searched,
where, and when. Concretely, the tooling's registry sweeps (Crossref,
OpenAlex, arXiv, DataCite) are logged activities with their query
terms, so ``no prior work found'' stops being an unfalsifiable
assertion and becomes a bounded one: no prior work found \emph{within
this recorded horizon}, which a sceptical reader can re-run, widen,
and possibly refute. The literature sweeps behind this paper are
logged exactly so. Every claim must then be supported, or refuted, by
facts, and both outcomes leave a record.

aiprov's answer makes claims first-class. Each assertion carries
\texttt{prov:wasGeneratedBy}, \texttt{prov:wasAttributedTo}, and a
ladder rung that separates ``the reference exists''
(machine-checkable) from ``the content supports the claim''
(machine-judged at most) from ``a human checked'' (human-only).
Refutation is explicit: \texttt{aiprov:contradicts} records a claim
refuted by facts as refuted, instead of silently deleting it. This is
supply-side transparency. Artefacts carry a pedigree that fabrication
cannot cheaply forge: DOIs that resolve, content hashes, commits, and
registry-resolved identities. Filtering the pool can then shift from
fluency heuristics to verifiable metadata.

\section{The aiprov Method}\label{sec:method}

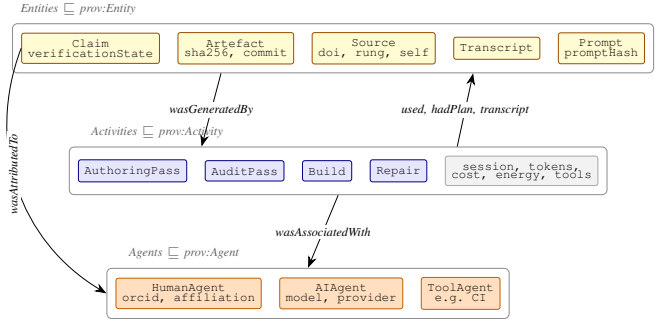
\begin{figure}[t]
\centering
\resizebox{\columnwidth}{!}{%
\begin{tikzpicture}[
  font=\scriptsize, node distance=2.5mm,
  ecls/.style={draw=orange!60!black, fill=yellow!22, rounded corners=1pt,
               inner sep=2.5pt, font=\tiny\ttfamily, align=center},
  acls/.style={draw=blue!50!black, fill=blue!10, rounded corners=1pt,
               inner sep=2.5pt, font=\tiny\ttfamily, align=center},
  gcls/.style={draw=orange!75!black, fill=orange!25, rounded corners=1pt,
               inner sep=2.5pt, font=\tiny\ttfamily, align=center},
  pkg/.style={draw=black!45, rounded corners=2pt, inner sep=3.5pt},
  pkl/.style={font=\tiny\itshape, text=black!60},
  e/.style={-{Stealth[length=1.7mm]}},
  lbl/.style={font=\tiny\itshape, fill=white, inner sep=1pt}]
\node[ecls] (claim) at (0,0)
  {Claim\\[-2pt]\tiny verificationState};
\node[ecls, right=of claim] (artefact)
  {Artefact\\[-2pt]\tiny sha256, commit};
\node[ecls, right=of artefact] (source)
  {Source\\[-2pt]\tiny doi, rung, self};
\node[ecls, right=of source] (transcript)
  {Transcript};
\node[ecls, right=of transcript] (prompt)
  {Prompt\\[-2pt]\tiny promptHash};
\node[pkg, fit=(claim)(prompt)] (epkg) {};
\node[pkl, anchor=south west] at (epkg.north west)
  {Entities \(\sqsubseteq\) prov:Entity};
\node[acls] (auth) at (0.6,-1.75) {AuthoringPass};
\node[acls, right=of auth] (audit) {AuditPass};
\node[acls, right=of audit] (build) {Build};
\node[acls, right=of build] (repair) {Repair};
\node[acls, right=of repair, draw=black!35, fill=black!5]
  (tele) {\tiny session, tokens,\\[-3pt]\tiny cost, energy, tools};
\node[pkg, fit=(auth)(tele)] (apkg) {};
\node[pkl, anchor=south west] at ([xshift=2mm]apkg.north west)
  {Activities \(\sqsubseteq\) prov:Activity};
\node[gcls] (human) at (1.4,-3.5) {HumanAgent\\[-2pt]\tiny orcid, affiliation};
\node[gcls, right=of human] (ai) {AIAgent\\[-2pt]\tiny model, provider};
\node[gcls, right=of ai] (toolag) {ToolAgent\\[-2pt]\tiny e.g.\ CI};
\node[pkg, fit=(human)(toolag)] (gpkg) {};
\node[pkl, anchor=south west] at ([xshift=2mm]gpkg.north west)
  {Agents \(\sqsubseteq\) prov:Agent};
\draw[e] ([xshift=-16mm]epkg.south) -- ([xshift=-19.5mm]apkg.north)
  node[lbl, midway] {wasGeneratedBy};
\draw[e] ([xshift=17mm]apkg.north) -- ([xshift=20mm]epkg.south)
  node[lbl, midway] {used, hadPlan, transcript};
\draw[e] (apkg.south) -- (gpkg.north)
  node[lbl, midway] {wasAssociatedWith};
\draw[e] (claim.west) to[bend right=40]
  node[lbl, pos=0.45, rotate=90] {wasAttributedTo} (gpkg.west);
\end{tikzpicture}%
}
\caption{Core aiprov vocabulary as an extension of PROV-O. Every class
subclasses a \texttt{prov:} term; tiny annotations name characteristic
attributes. Claims point to their generating activity, their
attributed agent, and a verification state; activities carry inference
telemetry and link prompts (as plans) and transcripts.}
\label{fig:vocab}
\end{figure}

\textbf{Model.} aiprov extends PROV-O with three agent classes
(\texttt{HumanAgent}, \texttt{AIAgent}, \texttt{ToolAgent}), four
activity classes (\texttt{AuthoringPass}, \texttt{AuditPass},
\texttt{Build}, \texttt{Repair}), and entity classes for artefacts,
claims, sources, transcripts, and prompts, where prompts are treated
as source code and bound to activities as plans
(Figure~\ref{fig:vocab}). Humans carry ORCID and affiliation; AI
agents carry model, version, provider, endpoint, context window, and
knowledge cutoff; tool agents cover deterministic actors such as CI
jobs. Treating prompts as plans deserves a word: a prompt is to an AI
activity what a script is to a batch job, the executable
specification of intent, so it is hashed and bound with
\texttt{prov:hadPlan} rather than paraphrased into a comment. An
auditor who wants to know \emph{why} an artefact came out as it did
reads the plan and the transcript; neither is recoverable after the
fact if not captured when the activity runs.

\textbf{Telemetry.} Activities carry session and request identifiers,
token counts by class (input, output, cache, reasoning), sampling
parameters, cost, energy, and the tools invoked. One rule governs all
of it: omit, don't estimate. Absent telemetry is recorded as absent;
invented precision is worse than a visible gap. Generated artefacts
receive sha256 content hashes at logging time, and prompt files
receive prompt hashes.

\textbf{Invariant I: no parentless claim.} Every \texttt{aiprov:Claim}
needs a generating activity, an attributed agent, and a verification
state. The conformance validator treats violations as hard failures
and is suitable as a pre-commit hook or CI gate.

\begin{figure}[t]
\centering
\resizebox{0.80\columnwidth}{!}{%
\begin{tikzpicture}[
  rung/.style={draw, rounded corners=1pt, minimum width=5.6cm,
               minimum height=0.52cm, font=\footnotesize, align=center},
  human/.style={rung, fill=black!12},
  note/.style={font=\scriptsize\itshape, anchor=west}]
\node[human] (r6) {6\; human-read};
\node[human, below=0.16cm of r6] (r5) {5\; human-confirmed};
\node[rung, below=0.16cm of r5] (r4) {4\; source-vendored};
\node[rung, below=0.16cm of r4] (r3) {3\; ai-confirmed};
\node[rung, below=0.16cm of r3] (r2) {2\; reference-resolved};
\node[rung, below=0.16cm of r2] (r1) {1\; needs-research};
\node[rung, below=0.16cm of r1] (r0) {0\; unverified};
\node[note, right=0.15cm of r6] {human-only};
\node[note, right=0.15cm of r5] {human-only};
\node[note, right=0.15cm of r4] {access gate for 5--6};
\node[note, right=0.15cm of r3] {highest AI-grantable};
\draw[-{Stealth}, thick] ([xshift=-0.5cm]r0.west) -- ([xshift=-0.5cm]r6.west)
  node[midway, sloped, above, font=\scriptsize] {promotions strictly climb};
\end{tikzpicture}%
}
\caption{The verification ladder. Every promotion is logged as an audit
activity; rungs 5--6 may never be granted by an AI agent, and the
validator treats such grants as hard failures.}
\label{fig:ladder}
\end{figure}
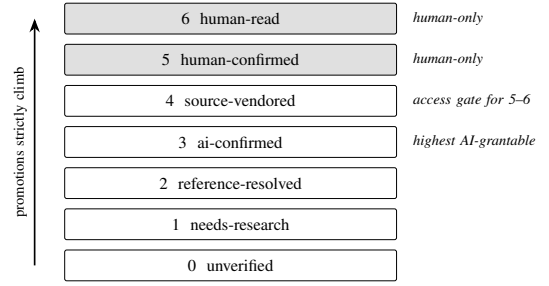

\textbf{Invariant II: the verification ladder.} Rungs answer who
checked what (Figure~\ref{fig:ladder}, Table~\ref{tab:ladder}).
Promotions must strictly climb and are themselves logged as audit
activities, so every rung a claim or source holds has a recorded
grantor. Disagreement is first-class: an agent may \emph{refuse} a
promotion, which changes no state but is logged as an audit activity
carrying the refused rung, so ``checked and not convinced'' is
distinguishable from ``never checked''. Rungs 5--6 are human-only; an AI-granted promotion to them is
a validator hard failure. Two rungs deserve comment. Vendoring
(rung~4) is not a verification statement: an unread vendored copy
proves nothing beyond \texttt{reference-resolved}. Its role is
instrumental, the access gate that makes human verification feasible
and durable. Operationally, the agent must hand the human the
evidence: \texttt{promote} refuses the human-only rungs unless the
source is vendored (path and hash recorded) or carries a clear access
link, and prints the review material with every request; a human
confirmation against link-only material is flagged, since the audit
target remains mutable. At the top, \texttt{human-read} subsumes
\texttt{human-confirmed}: it asserts full-text familiarity \emph{and}
confirmation, because a name alone would mislead, as reading does not
entail checking. Both refinements were operator-contested
mid-experiment (Section~\ref{sec:meta}). The ladder itself has
provenance: it canonicalizes labels field-tested in
\emph{Obscurity-Is-Dead}~\cite{obscurity-is-dead} and hardened in
F(AI)\textsuperscript{2}R~\cite{fair2r-repo}, and legacy names remain
machine-readable aliases, so that ancestor graphs can consolidate
onto this ladder without rewriting history; the aliases are in place,
though the consolidation has not yet been exercised against the
ancestor repositories' graphs.

\textbf{Commit binding.} An activity can only bind a commit that
already exists, and a commit cannot contain the graph entry describing
itself. The working discipline that follows: commit the artefacts, log
the activity with that commit's hash, then commit the graph update
separately. The graph never claims a hash it could not have known; the
cost is one extra commit per logged round. What the two commits buy
is tamper evidence in both directions. The graph entry names a commit
whose tree contains the artefacts it describes, content-hashed; the
graph itself lives in later commits of the same history. Backdating a
record would require rewriting published git history, and quietly
swapping an artefact would break its recorded hash: either
falsification is possible, but neither is silent; that asymmetry
(honest work is cheap, dishonest work is loud) is the design goal
throughout.

\textbf{The validator, exhaustively.} Conformance is seven concrete
checks, each a SPARQL query over the graph, and naming them makes the
invariants operational rather than aspirational. Three enforce
Invariant~I: no claim without a generating activity, no claim without
an attributed agent, and, as a warning, no agentless activity. Two
enforce Invariant~II: no human-only rung granted by an AI agent, and
no human-only rung without a recorded promotion activity naming a
human grantor, a warning that flags a rung edited directly
into the graph without an audit trail. Two enforce the access gate:
no source above \texttt{needs-research} without a resolvable DOI, URL,
or vendored copy, and a warning, rather than a failure, for
human-verified sources whose audit target is link-only and therefore
mutable. Four checks fail hard with a nonzero exit code;
that exit code is the entire integration contract, which is why the
same command serves unchanged as a pre-commit hook, a CI gate, and an
auditor's first step.

\textbf{Auditability operations.} \texttt{report} aggregates tokens,
cost, and rung distributions; \texttt{extract} computes the backward
closure of everything that contributed to chosen entities;
\texttt{dashboard} renders the graph as a self-contained offline
ledger. Sources enter through open registries (Crossref, OpenAlex,
arXiv, DataCite): a search command sweeps them, and registration with
DOI verification promotes a source to \texttt{reference-resolved} only
if the registry actually resolves it.

\begin{table*}[t]
\caption{The verification ladder. Positions are machine-readable
(\texttt{aiprov:ladderPosition}); promotions must strictly increase
them and are themselves logged as audit activities, so every rung a
claim or source holds has a recorded grantor.}
\label{tab:ladder}
\centering
\small
\begin{tabular}{clp{8.2cm}l}
\toprule
Pos & Rung & What it asserts & Who may grant \\
\midrule
0 & \texttt{unverified} & Recorded; nothing checked yet by anyone.
    Default state at creation. & (default) \\
1 & \texttt{needs-research} & A check was attempted or demanded and did
    \emph{not} succeed (e.g.\ a DOI failed to resolve). Must not be
    cited or relied upon until re-verified. & any agent \\
2 & \texttt{reference-resolved} & The reference \emph{exists}: its
    DOI/URL resolved in a bibliographic registry and metadata was
    captured. Says nothing about content. & any agent (tool-checkable) \\
3 & \texttt{ai-confirmed} & An AI checked that the source
    \emph{content} supports the associated claim. A machine judgement,
    the highest rung an AI may grant. & AI or human \\
4 & \texttt{source-vendored} & A copy of the source is preserved inside
    the repository (content-hashed), immune to link rot and silent
    revision. Of no evidential value by itself: this rung is the
    \emph{access gate} to human verification; the agent hands the
    human the evidence, and tooling refuses rungs 5--6 unless the
    source is vendored or carries a clear access link (DOI/URL). &
    any agent (hash-checkable) \\
5 & \texttt{human-confirmed} & A \emph{human} spot-checked that the
    source supports the claim. & human only \\
6 & \texttt{human-read} & A \emph{human} has read the source in full
    \emph{and}, with that full context, confirms the claim is
    supported, subsuming rung 5's confirmation with deeper
    familiarity. Reading without confirming is not a rung. Top of the
    ladder. & human only \\
\bottomrule
\end{tabular}
\end{table*}

\section{The Executable Skill}\label{sec:skill}

A method that lives in a PDF of guidelines depends on humans
remembering to follow it. We package aiprov instead as a
\emph{skill}: a versioned bundle of instructions and tools that an AI
agent loads and operates. The method's rules become the agent's
operating rules: log your own activities, record claims at
\texttt{ai-confirmed} at most, never fabricate telemetry, and never
grant a human-only rung.

\begin{figure}[t]
\centering
\resizebox{0.80\columnwidth}{!}{%
\begin{tikzpicture}[
  font=\scriptsize, node distance=2.2mm,
  stp/.style={draw=blue!50!black, fill=blue!10, rounded corners=1.5pt,
               inner sep=3pt, align=center, font=\scriptsize,
               minimum width=5.2cm},
  hstp/.style={stp, draw=orange!75!black, fill=orange!20},
  cstp/.style={stp, draw=black!50, fill=black!6},
  e/.style={-{Stealth[length=1.8mm]}},
  lbl/.style={font=\tiny\itshape, text=black!60, align=left}]
\node[hstp] (orcid) at (0,0) {the one setup question: ORCID iD};
\node[stp, below=of orcid] (init)
  {\texttt{init}: seed graph \(\cdot\) resolve identity \(\cdot\)
   scaffold CI + paper};
\node[stp, below=of init] (agent)
  {\texttt{agent}: register the AI system (model, provider)};
\node[stp, below=of agent] (work)
  {work round: author or edit artefacts \(\cdot\) commit};
\node[stp, below=of work] (log)
  {\texttt{log} activity binding that commit \(\cdot\) \texttt{claim}
   \(\cdot\) \texttt{source --verify}};
\node[stp, below=of log] (graphc)
  {commit graph \(\cdot\) \texttt{validate} \(\cdot\) push};
\node[cstp, below=of graphc] (ci)
  {CI (ToolAgent): validate \(\cdot\) dashboard \(\cdot\) disclosure
   \(\cdot\) PDF artefact};
\node[stp, below=of ci] (preview)
  {republish live preview at stable URL};
\node[hstp, below=of preview] (humanp)
  {human audit: \texttt{promote} to rungs 5--6 (gate hands over
   evidence)};
\foreach \a/\b in {orcid/init, init/agent, agent/work, work/log,
                   log/graphc, graphc/ci, ci/preview, preview/humanp}
  \draw[e] (\a) -- (\b);
\draw[e] (humanp.east) -- ++(0.45,0) |- (work.east)
  node[lbl, pos=0.25, rotate=90, anchor=south] {next round};
\end{tikzpicture}%
}
\caption{The skill's operating loop from the single setup question to
the continuously published build. Orange steps involve the human
operator; the grey step runs deterministically in CI; everything else
is the AI agent operating the method, two commits per logged round
(Section~\ref{sec:method}).}
\label{fig:workflow}
\end{figure}
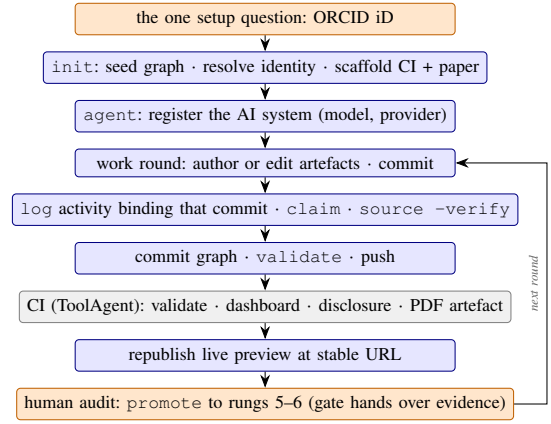

\textbf{ORCID-first setup.} Setup asks the human operator exactly one
question: their ORCID iD. A single command then seeds the graph,
resolves the operator's name and current affiliation from the public
registry (never guessed; if resolution fails, only the bare iD is
recorded), derives the instance IRI base from the git remote,
scaffolds continuous integration, and lays down a compilable
chapter-per-file paper skeleton whose author block is already filled
from the resolved identity. Later commands auto-detect the base from
the graph, so configuration never needs repeating
(Figure~\ref{fig:workflow}).

\textbf{CI as the deterministic build agent.} Every push validates the
graph server-side, renders the dashboard, regenerates the
AI-transparency disclosure, and compiles the paper to PDF, uploading
dashboard and PDF as build artefacts. The current version of the paper
is always available, and a conformance violation blocks the build. The
CI service itself is registered as a \texttt{ToolAgent}, and its green
runs are logged \texttt{Build} activities carrying artefact digests.
Version tags extend the same discipline to releases: a \texttt{v*}
tag publishes a conformance-gated release whose assets are the PDF
together with the graph it was built from, the metrics snapshot, the
packaged skill, and their checksums, so a release is the artefact
plus its record.

\textbf{Live preview as oversight affordance.} On request, the agent
republishes the current typeset build, together with the per-section
planning skeletons, to a stable preview URL after every editing round
(Figure~\ref{fig:environment}, right panel). Effective oversight of
a fast-moving AI collaborator needs a continuously current view of
the deliverable; the EU AI Act's human-oversight
expectation~\cite{eu-ai-act2024} becomes a user-interface property.
The preview also renders the source-verification queue, whose
buttons open prefilled repository issues; a workflow verifies the
issue author against a registered-operator mapping and transcribes
the authenticated grant or refusal into the graph, so the operator
walks the human rungs without touching a checkout: the judgement
stays the human's, only its transcription is automated.

\textbf{Structure before prose.} The agent first lays down each
section's principal structure, its argument flow, evidence, and figure
and citation plan, as a reviewable skeleton; prose is a separate later
pass. The operator can contest structure cheaply before any prose
exists to defend it, and both passes are logged, so the audit trail
shows which structural decisions preceded which text. Two rungs of the
ladder were redesigned exactly this way during this experiment.

\textbf{Regulatory fit by construction.} The EU AI Act transparency
statement is derived from the graph on every build
(Section~\ref{sec:discussion}); disclosure is a build product, not
an authoring chore, and adopters inherit the mechanism.

\textbf{Worked example: analysing experimental data.} This example
is a prospective design, not yet an implemented deployment. The
workflow is not specific to writing papers; consider a study like the comparison
of power and amplitude control in continuous ultrasonic welding of
unidirectional carbon-fibre-reinforced polymers
(CFRPs)~\cite{ultrasonic-welding2025}, where welding
trials produce in-situ process data, scans, mechanical test results,
and micrographs, and the outcome is comparative claims about two
control strategies. Setup is identical: one ORCID question seeds the
graph and registers the experimenters; the rig's data acquisition and
the analysis scripts join as \texttt{ToolAgent}s, the AI assistant as
an \texttt{AIAgent}. Each raw record (a weld's power trace, a scan,
a test protocol) is registered as a \texttt{Source} and
vendored with its content hash, so the access gate of rung~4 holds the
evidence for every downstream claim as immutable bytes. Each analysis
run is a logged \texttt{Build} activity that \texttt{used} the raw
sources and \texttt{generated} the derived tables and figures, hashes
included; CI reruns the analysis, so a figure that cannot be
regenerated from the recorded inputs fails the build. AI-assisted
interpretation is an \texttt{AuthoringPass} with telemetry, and every
finding (say, that the two control modes yield equivalent weld quality
on a given layup) is a claim generated by the analysis activity and
attributed to its agent. The ladder then works exactly as in
Section~\ref{sec:method}: the AI may cross-check a claim against the
derived data and grant \texttt{ai-confirmed}; the experimenter
spot-checks the mechanical-test table for \texttt{human-confirmed},
with the tooling handing over the vendored data on request; the
comparative conclusions that carry the paper deserve
\texttt{human-read}. The disclosure statement and the report come for
free, and once the shepard integration planned in
Section~\ref{sec:conclusion} lands, raw data and provenance will
share one backbone.

\textbf{Auditability at the command line.} Auditability here is not an
abstract property but a concrete session: an auditor clones the
repository and runs the same tooling the authors ran. Three checks
need no trust in anyone's honesty. Re-running the validator replays
every invariant against the graph as it stands; re-hashing a vendored
file or a committed artefact and comparing against the recorded digest
detects any silent swap; re-resolving a recorded DOI confirms the
citation metadata still matches the registry. Two further checks read
rather than execute: the activity log tells the auditor \emph{who}
granted each rung and under which commit, and the transcript shows the
dialogue in which a contested decision was made. The tooling is built
so that success is quiet and failure is loud,
Figure~\ref{fig:cli}: a clean graph prints one \texttt{OK} line per
invariant and exits zero, suitable as a CI gate; a violation names the
offending node and exits nonzero; a refused operation says why it was
refused and what would make it legitimate. The last property matters
most for the human rungs: when the operator asks for
\texttt{human-confirmed}, the gate either blocks for lack of access
material or prints exactly where the review copy sits, so the human
step starts with the evidence in hand rather than a search for it.

\begin{figure}[t]
\centering
\fontsize{6.6}{7.6}\selectfont
\begin{verbatim}
$ provlog.py validate
[ OK ] parentless claims: 0
[ OK ] unattributed claims: 0
[ OK ] human-only rungs granted by AI agents: 0
    (...four further checks OK...)        exit 0

# an AI agent tries to grant a human-only rung
$ provlog.py promote --id demo-claim \
    --to human-confirmed --agent claude-code
REFUSED: rung 'human-confirmed' is human-only;
  agent:claude-code is not a registered
  HumanAgent. An AI must never grant this rung.
                                          exit 1
# a parentless claim spliced into the graph
[FAIL] parentless claims: 1
    -> .../prov/claim/rogue               exit 1
# access gate: human rung, source not in hand
$ provlog.py promote --id demo-src \
    --to human-confirmed --agent florian-krebs
REFUSED: 'human-confirmed' requires the source
  in hand -- vendor a copy first (promote
  --to source-vendored --file <path>) or
  record a DOI/URL for it                 exit 1
# vendor the copy, then the gate hands it over
$ provlog.py promote --id demo-src \
    --to source-vendored --file protocol.pdf
demo-src: reference-resolved -> source-vendored
$ provlog.py promote --id demo-src \
    --to human-confirmed --agent florian-krebs
review material: vendored copy at protocol.pdf
demo-src: source-vendored -> human-confirmed
  (logged as AuditPass, agent:florian-krebs)
                                          exit 0
\end{verbatim}
\caption{Success and failure, captured from the tooling running
against a sandbox copy of this paper's graph: a clean validation, the
two invariant refusals (human-only rung attempted by an AI; parentless
claim), and the rung-4 access gate first blocking a human promotion
and then, once a copy is vendored, handing the review material to the
human. Output verbatim; long lines re-wrapped and exit codes
annotated.}
\label{fig:cli}
\end{figure}

\textbf{Versioning the method.} The skill tree is the repository:
snapshots are immutable, changes are logged, and every activity binds
its commit, so the meta-experiment can cite the exact method version
in force at any point. The same property makes the skill the
transferable unit; the bundle that wrote this paper is the
intended integration payload for the shepard fork of
Section~\ref{sec:conclusion}. Conversations are part of the record
too: a transcript exporter renders the session to a committed file
after every turn, and activities link to it, so the dialogue that
shaped an artefact is as retrievable as the artefact.
\ifextended

\textbf{Designed extensions.} Three tool-level extensions are
specified against the same invariants and are reported here as
designs, not shipped features, in keeping with the rule that the
record must not outrun the tooling. First, a \emph{distributed
reviewer library}: a verification event is already self-contained,
source identifier, rung, grantor, evidence hash, so it can be
exported as a signed, portable record and pooled into personal and
shared registries. A project importing such a registry sees a
source's verification history across projects, with every rung still
bound to its original grantor; human rungs are never inherited
automatically but either re-granted locally or imported as attributed
third-party audits, the third-witness pattern of
Section~\ref{sec:discussion}. Second, \emph{backward closure with
citation binding}: because every claim carries its parent activity
and sources, a state change on a source, a refutation or retraction,
can be propagated backward mechanically to enumerate every claim,
citing sentence, and downstream artefact resting on it. Third, a
\emph{dispute-ready workflow}: \texttt{aiprov:contradicts} already
records refutation, and the extension routes disputes through the
same authenticated issue path as promotions, so contesting a claim
produces a structured, attributed record, who disputes what on which
evidence, instead of a comment thread.
\fi

\section{Meta-Experiment: This Paper as Case Study}\label{sec:meta}

\begin{figure*}[t]
\centering
\includegraphics[width=0.88\textwidth]{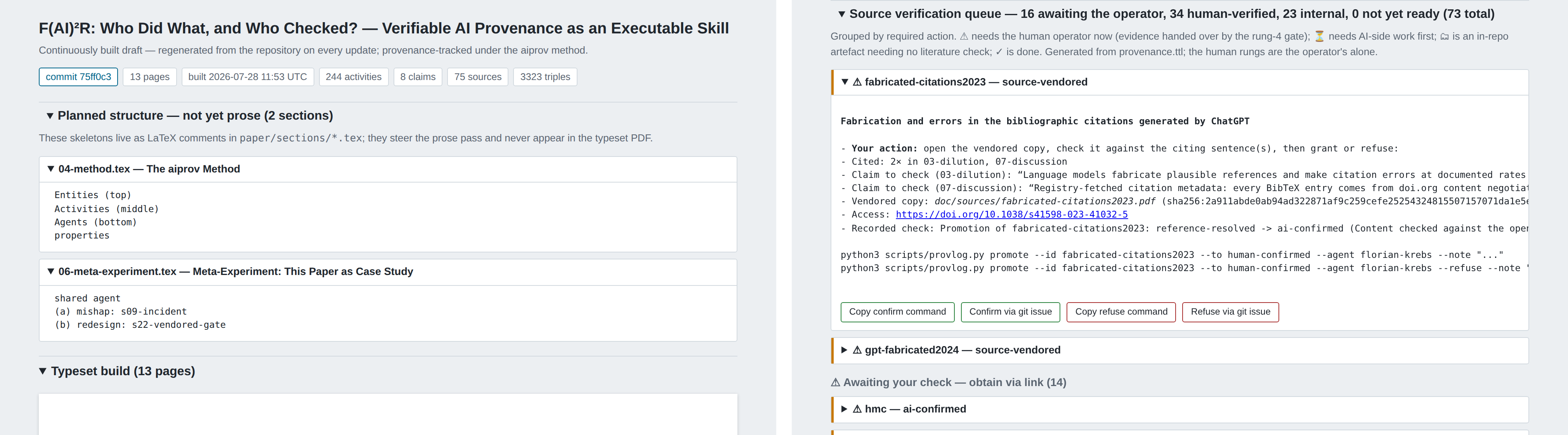}
\caption{The continuously rebuilt preview of this paper (screenshot,
2026-07-28), republished to a stable URL after every editing round.
Left: commit, build time, and live graph statistics above the typeset
pages. Right: the source-verification queue at the bottom of the same
page; each entry shows the citing sentence to check, the recorded
evidence, the access link, and buttons that stage the exact grant or
refuse command or open a prefilled repository issue, which a workflow
transcribes into the graph after verifying the issue author against
the registered-operator mapping. The judgement stays the operator's;
only its transcription is automated.}
\label{fig:environment}
\end{figure*}

\textbf{Setup.} The repository was bootstrapped by an AI agent
operating the skill, in one working session directed by the human
operator. Three agents are registered: one human (identity resolved
from ORCID), one AI system, and the CI service as a tool agent. Every
round of work follows the two-commit pattern of
Section~\ref{sec:method}, so each logged activity binds a real commit.
Figure~\ref{fig:environment} shows the oversight surface: the
continuously rebuilt draft beside its source-verification queue, the
loop of Section~\ref{sec:skill} in operation. Scope, before any number:
this is a demonstration, not a study; one operator, one domain, one
session. The quantities below characterize this instance of an idea
in use (Section~\ref{sec:conclusion} lists the trials it owes).

\begin{figure}[t]
\centering
\includegraphics[width=\columnwidth]{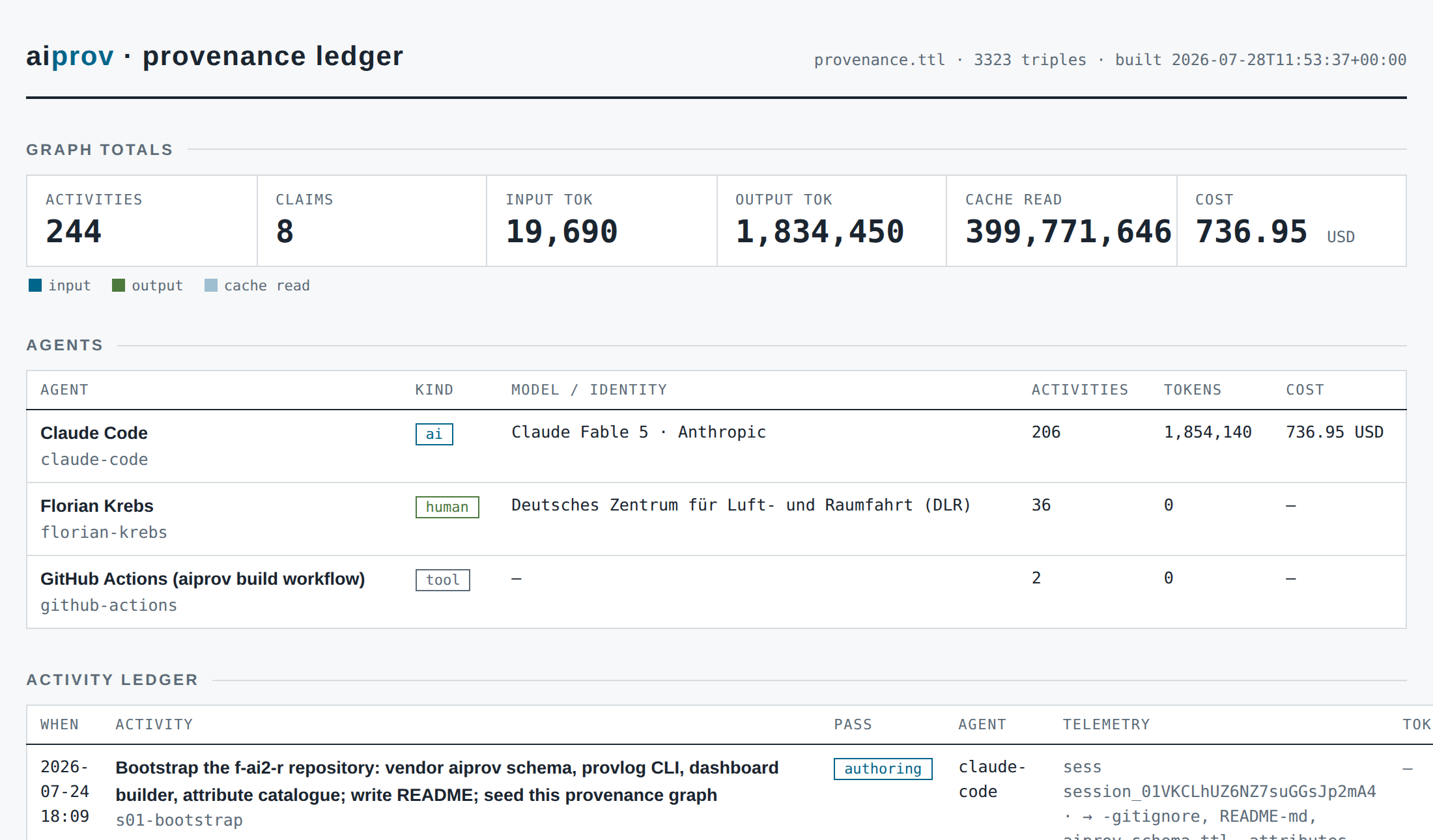}\\[2pt]
\includegraphics[width=\columnwidth]{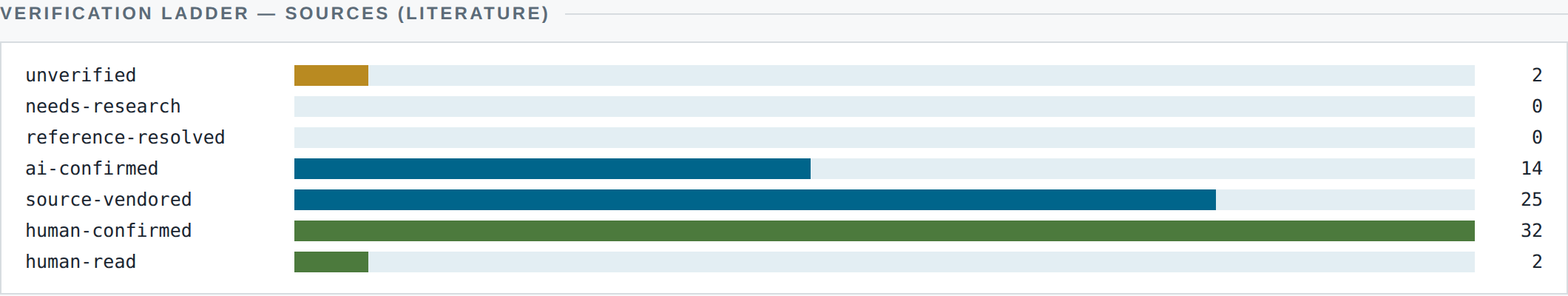}
\caption{The dashboard rendered from \texttt{provenance.ttl}
(screenshot, 2026-07-28). Top: header, graph totals, agents, and the
start of the activity ledger; the token and cost tiles sum
per-activity records (two aggregate backfills, roughly half the
computed cost), so the session-record aggregates of
Table~\ref{tab:telemetry} are the authoritative figures. Bottom: the literature rung distribution, most sources
operator-verified, the top rung populated. The full page is one
self-contained HTML file with an interactive graph view.}
\label{fig:dashboard}
\end{figure}

\textbf{The graph as evaluation object.} Table~\ref{tab:telemetry}
reports the state of the graph at the time of writing, and
Figure~\ref{fig:dashboard} shows the ledger view rendered from the
same file. Token counts
were initially absent: the harness (the agent's execution
environment) exposes no live usage to the agent,
and under omit-don't-estimate nothing was recorded. The operator's
challenge (``why is this not reported?'') prompted a check of the
session record, which turned out to carry the provider's per-request
usage blocks; the totals in the table are aggregated from them,
reported not estimated. Cost followed one step behind: the record
carries no priced entries, so cost was first omitted; the
operator then directed pricing the recorded tokens
against the provider's published price list at time of writing. The
resulting figure is \emph{computed, not provider-reported}, and the
graph records it exactly so, price basis in the activity's label. All
eight recorded claims sit at \texttt{ai-confirmed}, the ceiling an AI
may grant; promotion beyond waits for the human operator, with the
evidence handed over on request.

\begin{table}[t]
\caption{Provenance-graph state of this paper's production at the time
of writing (from \texttt{provlog.py report} and the repository;
regenerated before submission). Token totals are aggregated from the
provider-reported usage blocks in the session record; the cost is
computed from those tokens and the published price list at time of
writing (all cache writes in the record are 1-hour-TTL), not reported
by the provider, and is recorded in the graph as computed.}
\label{tab:telemetry}
\centering
\begin{tabular}{lr}
\toprule
Metric & Value \\
\midrule
Triples & \MTriples \\
Logged activities & \MActivities \\
Claims (all \texttt{ai-confirmed}) & \MClaims \\
Sources on the ladder & \MSources \\
\quad thereof at \texttt{ai-confirmed} or above & \MSrcChecked \\
\quad thereof marked self-citation & \MSrcSelf \\
Registered agents (human / AI / tool) & 1 / 1 / 1 \\
Commits & \MCommits \\
Transcript & 1 session, \MTurns{} turns \\
Output tokens & \MOutTok \\
Input tokens (uncached) & \MInTok \\
Cache tokens (read / write) & \MCacheReadM\,M / \MCacheWriteM\,M \\
Cost (computed, not reported)$^{\dagger}$ & $\approx$ USD~\MCostUSD \\
\quad same tokens at Opus~5 / Sonnet~5 rates$^{\dagger}$ &
  $\approx$ USD~\MCostOpus{} / \MCostSonnet \\
\bottomrule
\end{tabular}
\par\smallskip
{\raggedright\footnotesize $^{\dagger}$Basis: the provider's published
API list prices~\cite{anthropic-pricing}, per million tokens:
\MPriceBasis{}; all cache writes 1\,h-TTL. The repricing holds the
recorded token volume fixed and only swaps the price list (Sonnet~5
at introductory rates); it compares price bases, not runs, since
models tokenize and behave differently.\par}
\end{table}

\textbf{Worked examples.} Invariant enforcement was verified by
test-run before any claim entered the graph: an injected parentless
claim failed validation, and a promotion of a claim to
\texttt{human-confirmed} attempted by the AI agent was refused, both
captured in Figure~\ref{fig:cli}. The
bibliography was built through the method: sources found by registry
sweeps, DOI-verified at registration, fetched as BibTeX from doi.org
content negotiation, and cited only once ladder-backed. The paper's
intellectual ancestors are themselves registered sources, each
fetched or URL-verified first. Registration, however, only proves a
reference
\emph{exists}; an audit pass flagged exactly this as the record's
weakest point, and content passes closed it: for every cited source
the AI fetched the actual content (full text or slides where openly
accessible, abstracts otherwise) and checked it against the specific
citing sentence, recording the supporting quotation and the checked
depth in the promotion note; every cited source was checked at
the time of writing. Negative outcomes stayed recorded rather than
smoothed over: one publisher's bot-wall deferred a check until an
open index carried the abstract, the operator-owned repositories were
left to the operator's own rungs, and three citation overreaches
surfaced by the checks were repaired in the text before any
promotion.

\begin{figure*}[t]
\centering
\begin{tikzpicture}[
  font=\scriptsize,
  act/.style={draw=blue!50!black, fill=blue!10, rounded corners=1.5pt,
              align=center, inner sep=3.5pt},
  ent/.style={draw=orange!60!black, fill=yellow!22, ellipse,
              align=center, inner sep=1.5pt},
  clm/.style={ent, very thick},
  ag/.style={draw=orange!75!black, fill=orange!22, regular polygon,
             regular polygon sides=5, inner sep=1.5pt, align=center},
  st/.style={draw=black!45, fill=black!5, align=center, inner sep=2.5pt},
  attr/.style={draw=black!35, fill=black!4, align=left, inner sep=3pt,
               font=\tiny},
  lbl/.style={font=\tiny\itshape, fill=white, inner sep=1pt},
  e/.style={-{Stealth[length=1.8mm]}}]
\node[ag] (agent) at (0,0) {agent:\\claude-code\\\tiny AIAgent};
\node[act] (s09) at (-4.6,0) {act:s09-incident\\\tiny aiprov:AuditPass};
\node[attr] (s09a) at (-4.6,-1.9) {endedAtTime 2026-07-24T18:45Z\\
sessionId session\_01VKC\ldots\\ usedTool bash, git\\
label: ``\ldots overwrote two committed\\
sections \ldots restored from 75d21bd''};
\draw[e] (s09) -- node[lbl]{wasAssociatedWith} (agent);
\draw[black!35, dotted] (s09) -- (s09a);
\node[act] (s22) at (4.6,0) {act:s22-vendored-gate\\\tiny aiprov:AuthoringPass};
\draw[e] (s22) -- node[lbl]{wasAssociatedWith} (agent);
\node[ent] (e1) at (2.9,-1.8) {provlog.py\\\tiny Artefact + hash};
\node[ent] (e2) at (5.1,-2.1) {aiprov-schema.ttl\\\tiny Artefact + hash};
\node[ent] (e3) at (7.1,-1.5) {SKILL.md\\\tiny Artefact + hash};
\draw[e] (e1) -- node[lbl]{wasGeneratedBy} (s22);
\draw[e] (e2) -- (s22);
\draw[e] (e3) -- (s22);
\node[clm] (c7) at (1.7,1.9) {claim:c7\\\tiny ``promote refuses human rungs\\\tiny without vendored copy or link''};
\node[st] (vs) at (5.6,2.5) {verif:ai-confirmed};
\draw[e] (c7) -- node[lbl]{wasGeneratedBy} (s22);
\draw[e] (c7) to[bend right=12] node[lbl, pos=0.4]{wasAttributedTo} (agent);
\draw[e] (c7) -- node[lbl]{verificationState} (vs);
\end{tikzpicture}
\caption{Two incidents from this experiment, exactly as the graph
represents them (node contents abridged from real triples). Left, a
mishap: the agent overwrote two committed manuscript sections; the
self-report is an \texttt{AuditPass} carrying timestamp, session,
tools, and the restoring commit in its label. Right, a contested
design: the operator challenged the \texttt{source-vendored} rung's
semantics; the redesign activity generated new tool and schema
versions (content-hashed artefacts), and its enforcement claim sits at
\texttt{ai-confirmed} after test-run. The incident vocabulary is just
the ordinary vocabulary.}
\label{fig:incident}
\end{figure*}
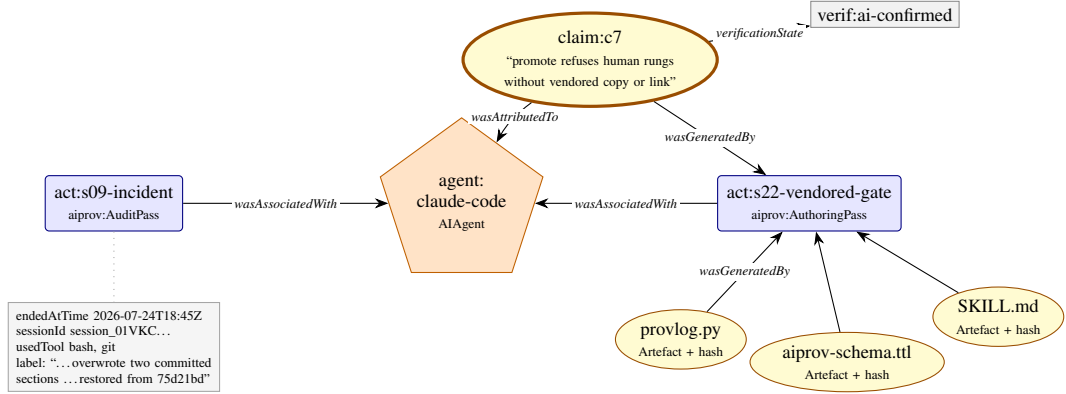

\textbf{Incidents as first-class records.} The experiment produced
four incident types, and none needed special vocabulary; each is an
ordinary activity whose class carries its nature
(Figure~\ref{fig:incident}). (i)~An agent mishap: while creating
template stubs, a stale working directory led the agent to overwrite
two committed manuscript sections. Version control surfaced the
unintended diff immediately, restoration was a one-command revert, and
the self-report is an \texttt{AuditPass} naming what happened, the
tools involved, and the restoring commit. (ii)~An infrastructure
failure: a registry served malformed BibTeX, which broke the CI build;
the diagnosis and fix live in commits, and the first green run is a
logged \texttt{Build} activity by the CI tool agent carrying artefact
digests. (iii)~Blocked access: a publisher returned 403 to automated
requests, and the fallback to the canonical page is recorded in the
registering activity's label rather than silently swallowed. (iv)~Method
contestation: the operator challenged the semantics of two ladder
rungs; the redesigns are ordinary authoring activities whose generated
artefacts are the new tool and schema versions, with the enforcement
claim verified by test-run. The pattern generalizes: erasing an
incident would require rewriting git history \emph{and} the graph, both
visible acts, so honesty is the cheap path. The same property that
makes contributions attributable makes errors reversible.

\textbf{Division of labour as data.} The record makes the division of
work between operator, AI agent, and infrastructure countable
(Table~\ref{tab:division}). The class-by-agent counts are exact,
machine-readable queries over the graph; attributing \emph{triggers},
what prompted each activity, still requires reading the session log,
and that reading is consistent: the operator's direction messages
carry scope, lineage, people, and every contested semantic decision
(both ladder redesigns, the cost policy, the license choice, the
page budget), while the AI executed all authoring, audit, and repair
activities and CI contributed the deterministic builds. Rungs above the AI ceiling remain human by
design: the operator granted the first \texttt{human-confirmed}
against a vendored statute text the gate handed over, confirmed
the five self-cited sources they authored, the class where the
human's judgement is uniquely authoritative, and then walked the
queue through the issue path of Section~\ref{sec:skill},
\MHumanRungs{} human rungs granted at the time of writing.
Authentication splits those grants honestly: \MGrantsIssue{}
promotions carry GitHub-authenticated authorship through the issue
path, itself derived mid-experiment, while \MGrantsSession{} were
transcribed in session from the operator's instruction, checkable
only against the committed transcript; \MLinkOnly{} of the
human-verified sources remain link-only audit targets, the
validator's warning kept visible. The
remaining sources are deliberately kept at their machine rungs so
every rung was exercised over the record's history
(the final distribution occupies the upper four); the
operator reports having checked all sources and claims, and the
ladder records which of those checks they chose to formalize as
grants. The hypothesis that AI absorbs tedium while humans
connect ideas and people (Section~\ref{sec:discussion}) is thereby
testable from the committed log and transcript, for this one
experiment. One measurement pitfall became a method fix: the session
format returns tool results under the \emph{user} role, so a naive
role count inflates the human by an order of magnitude (595 of 663
user-role messages were tool results). The transcript exporter now
labels Human, Assistant, Tool result, and Hook distinctly; the same
unambiguous role labeling is worth asking of provider session formats
generally, in the same breath as provider-reported cost.

\textbf{Leverage as data.} The same record measures what the
multiplication bought, without estimating any counterfactual. The
operator typed \MDirWords{} words of direction in total, median
\MDirMedian{} words per message; active session time, clustering
request timestamps and counting gaps under thirty minutes, is about
\MActiveH{} hours inside a wall-clock span of \MSpanH{} hours. Against
that input stand roughly \MProseWordsK{} words of typeset prose,
about \MToolLinesK{} lines of tooling, \MSources{} ladder-backed
sources, and the graph itself. What a solo baseline would have cost
is deliberately not estimated, omit-don't-estimate applies to
counterfactuals too; what the record does show is \emph{where} the
freed attention went: nearly every typed word carries scope,
semantics, lineage, or people, not prose. The gain is less the
saved time than this: the human's whole contribution could be
judgement. Every derived figure above follows the skill's
derivation rule: its methodology is stated with the number, word counts strip LaTeX markup, direction counts exclude tool results,
hook messages, skill loads, and pasted documents over five hundred
words, and the thresholds ride in the measuring activity's label, so
a sceptic can recompute or contest any of them.

\begin{table}[t]
\caption{Division of work in the meta-experiment, counted from the
provenance graph and the session record at the time of writing
(regenerated before submission).}
\label{tab:division}
\centering
\begin{tabular}{p{5.2cm}lr}
\toprule
Contribution & Agent & Value \\
\midrule
Direction messages: scope, lineage, people, contested semantics (two
ladder redesigns, cost policy, license, page budget) & human & \MDirMsgs \\
Typed direction volume (median \MDirMedian{} words/message) & human & \MDirWords{} w \\
Active session time (request-gap clustering) & both & \MActiveH{} h \\
Authoring passes: prose, code, figures, method releases & AI & \MPassAuth \\
Audit passes: validation, promotions, content checks, incident
reports, measurements & AI & \MPassAudit \\
Repair passes & AI & \MPassRepair \\
Deterministic builds logged & CI & \MPassBuild \\
Rungs granted above the AI ceiling & human & \MHumanRungs \\
\bottomrule
\end{tabular}
\end{table}

\textbf{What the auditor can and cannot verify.} Hashes, commits,
resolving DOIs, rung grants, and the validator's verdicts are
checkable by anyone with the repository. The honesty of self-reported
telemetry is not; that limit is where the discussion begins.

\section{Discussion and Limitations}\label{sec:discussion}

\textbf{Self-report honesty.} The graph is only as truthful as its
writer. The mitigations are structural: content
hashes and commit bindings tie records to bytes, validation runs
server-side in CI, the omit-don't-estimate rule removes the temptation
of invented precision, and the human-only rungs cap what an AI may
assert about its own work. The overwrite incident of
Section~\ref{sec:meta} shows the mitigation working: with the
repository as the process substrate, an AI mistake was a recoverable,
recorded event rather than silent corruption.

\textbf{From honesty to evidence: the dual-witness ask.} The
remaining weakness is that the record has one witness. Telemetry is
provider-generated but agent-relayed: nothing stops a dishonest agent
from underreporting usage, omitting sessions, or inventing
activities. Provider-side provenance, signed inference records
carrying model, timestamps, usage, roles, and prompt/response
hashes, would eliminate that class of misreporting and subsume this
paper's other provider asks (cost, energy, unambiguous roles). It would not, however, be
tamper-\emph{proof}, on three grounds. Trust shifts rather than
vanishes: a signature proves the provider asserted the record, so
key custody moves to the centre. Coverage stops at the API seam: the
server can attest what passed through it, not which files were
touched or how an attested inference binds to a repository artefact,
so a dishonest client can still mis-bind. And judgements stay
judgements: a signature proves a model emitted a verdict, not that
the verdict is sound, and no server can attest that a human read a
source. The
realistic endpoint is a \emph{dual witness}: provider-signed
inference logs and repository-side commit binding, each referencing
the other's hashes, so that falsifying the record requires collusion
between independent parties with separately published histories, the
same trust structure as certificate transparency or double-entry
bookkeeping. A third witness is nearly free: anchoring each release's
checksums in an append-only public log (certificate-transparency
style, or a distributed ledger) makes retroactive rewriting of the
repository history detectable without trusting its host, though
anchoring, like signing, certifies existence, never truth. aiprov
supplies one witness and the seam for the others; signed provenance
is the strongest standing ask this paper directs at providers.

\textbf{Why human-only rungs stay human.} \texttt{ai-confirmed} is a
machine judgement with known failure
modes~\cite{ai-hallucination2023}. The ladder does not pretend
otherwise; it makes the judgement's author explicit instead of
laundering machine confidence into apparent human endorsement. The
human rung is no formality either: even domain experts struggle to
override erroneous AI judgements against contradicting
evidence~\cite{doctors-vs-algorithms2026}; hence the rung-4 gate
hands the operator source and citing sentence, evidence before
judgement.

\begin{figure}[t]
\centering
\begin{tikzpicture}[x=0.0058\columnwidth, y=0.30cm, font=\scriptsize]
\foreach [count=\i] \v/\l in {\MOverReqNum/{provider requests},
    \MOverOutNum/{output tokens}, \MOverCostNum/{computed cost},
    \MBookCommitsPctNum/{commits}} {
  \fill[black!10] (0,-\i) rectangle (100,-\i+0.62);
  \fill[black!60] (0,-\i) rectangle (\v,-\i+0.62);
  \node[left] at (-1,-\i+0.31) {\l};
  \node[right] at (\v,-\i+0.31) {\v\,\%};
}
\draw (0,-4.55) -- (100,-4.55);
\foreach \t in {0,25,50,75,100}
  \draw (\t,-4.55) -- (\t,-4.75) node[below] {\t};
\end{tikzpicture}
\caption{Method overhead as shares of the recorded session and
repository, computed by the request classification described in the
text and drawn from the metrics single source. The commits bar
counts bookkeeping-only commits; every logging round commits,
however small, hence its larger share.}
\label{fig:overhead}
\end{figure}
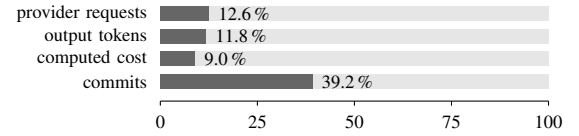

\textbf{Cost, and a gap that half closed.} The method's overhead in
this experiment was one logging command and one extra commit per
working round, plus the audit activities the tooling writes itself.
Because the session record carries per-request usage and tool calls,
that overhead is measurable rather than anecdotal: a mechanical
classification of every provider request by its tool
invocations, bookkeeping commands (logging, validation, promotion,
graph and transcript commits) versus everything else, attributes
\MOverReqPct{} of requests, \MOverOutPct{} of output tokens, and
\MOverCostPct{} of the computed session cost to the method itself;
on the repository side, \MBookCommits{} of \MCommits{} commits
carried only graph or transcript updates. No request mixed bookkeeping with substantive
work, so upper and lower bounds coincide: roughly a tenth of the
activity is the price of making the other nine tenths auditable
(Figure~\ref{fig:overhead}).

Token telemetry illustrates the omit-don't-estimate rule in motion:
the harness exposes no live usage to the agent, so nothing was
recorded at first; when the operator challenged the gap, the session
record turned out to carry the provider's per-request usage blocks,
and the totals in Table~\ref{tab:telemetry} were backfilled from
them, reported rather than estimated. Cost closed the same way, one
step later and one grade weaker: the record is unpriced, so under
omit-don't-estimate no cost was recorded until the operator directed
pricing the recorded tokens against the provider's published price
list, first roughly USD~415, since refreshed in
Table~\ref{tab:telemetry}. The graph preserves the distinction
between \emph{reported} telemetry (token counts from the provider's
usage blocks) and \emph{computed} telemetry (cost, derived from them
at a disclosed price list). Computed values are
honest only while their basis is recorded; provider-reported cost
would still be the stronger record. A larger question stays open behind the
numbers: this one session consumed \MCacheReadMint~million cache-read
tokens, USD~\MCostUSD{} in all at list prices, for one draft paper
and one method release, and whether that justifies its environmental
and other costs is a debate this paper does not settle and does not
hide. The schema reserves \texttt{aiprov:energyWh} for exactly this
reason; it sits empty because no provider reported energy. Our
position is only that the debate belongs over recorded consumption,
not guesses, in either direction: neither dismissing AI assistance
as obviously wasteful nor excusing it as obviously worthwhile. The
graph makes the bill legible; whether it is worth paying is a
judgement the record enables but cannot make.

\textbf{Skill drift.} The method evolved mid-experiment, v0 to \MSkillVersion{},
including two operator-contested ladder redesigns. Immutable archived
snapshots, a changelog, and per-activity commit binding keep every
run attributable to the exact method version in force, so evolution
does not blur attribution.

\textbf{Practices that emerged.} Four, each small enough to adopt
tomorrow; the first two are known from Manubot's continuous
manuscripts~\cite{manubot2019}, rediscovered here in an
agent-operated setting. Registry-fetched citation metadata: every BibTeX entry
comes from doi.org content negotiation, never typed from memory, a
concrete discipline against fabricated
references~\cite{fabricated-citations2023}; the one failure was a
registry serving malformed BibTeX, and CI caught it, not a reader.
Registry-resolved identities: people enter the record via ORCID
lookups, extending never-fabricate to social metadata.
Verified-before-cited: no reference without a ladder-backed source
node, making the bibliography a view of the graph. And
content-checked before load-bearing, born from this paper's own
audit (Section~\ref{sec:meta}): any citation an argument leans on
should reach \texttt{ai-confirmed} at minimum, with the checked
depth disclosed in the promotion note.

\textbf{Transparent self-citation.} No-parentless-claim forces citing
one's own priors; the lineage of Section~\ref{sec:background} is the
author's own repositories and specifications. Self-citation is
frowned upon as metric gaming, yet unavoidable as provenance. The
resolution is transparency instead of omission: sources carry
\texttt{aiprov:selfCitation}, the report prints the ratio (here \MSrcSelf{} of
\MSources), and a reviewer can judge necessity per source; all five
are operator-confirmed, the validator's link-only warning kept
visible. The test that
separates influence from inflation: a transparent self-citation is
load-bearing, removing it would orphan a lineage or method claim, and
it points at repositories and specifications rather than
metric-bearing venues. It declares descent; it does not farm
citations.

\textbf{Regulatory alignment.} The EU AI Act's transparency rules ask
for disclosure and machine-readable marking of AI-generated
content~\cite{eu-ai-act2024}. A provenance graph is exactly that
marking, and because CI derives the disclosure statement from the
graph on every build, the published statement cannot lag the record.
The Commission now ships free icons for marking content as
AI-generated or AI-modified under Article~50(4) and its Code of
Practice~\cite{eu-ai-icons2026}; the labelling duty for published
text falls away under human review with editorial responsibility,
which is precisely what the ladder records, so the graph
substantiates label and exemption alike. Research-integrity guidance
asks the same from the norms side: a FAQ from the German ombuds
context records consensus that AI use be declared, tool, version,
date, and manner, with the specifics still under-defined across
publishers~\cite{faq-ai-gwp2025}; a graph-derived disclosure meets
that common core mechanically.
Beyond the EU, the OECD AI Principles, the first intergovernmental
AI standard, ask AI actors to ``ensure traceability, including in
relation to datasets, processes and decisions made during the AI
system lifecycle''~\cite{oecd-ai-principles}: a per-activity
provenance graph in all but name, and the OECD.AI observatory's
incident monitoring~\cite{oecd-aim} aggregates exactly such
per-project records.
Provenance-first authoring turns a compliance duty into a
by-product.

\textbf{Generality.} Beyond papers, the same invariants and ladder
apply to code, data pipelines, CAD models, and AAS
submodels~\cite{aas-part1-v30rc02}; what remains domain work is the
choice of entity subclasses. Generality also holds across clients and
models: the skill bundle follows the open Agent Skills
format~\cite{agent-skills-spec}, which independent agent clients
implement, so the same bundle loads beyond the environment that
produced this paper, and the graph is provider-agnostic, since any AI
system, hosted or local, registers as an \texttt{AIAgent} with model
and provider as data. Only three capture shims are client-specific
(turn-end hook, transcript export, usage-field mapping); graph,
ladder, and validator are plain Python and git. The AAS case bridges
to the outlook: digital-twin infrastructures already assume
machine-readable semantics, so AI-work provenance slots in as one more
submodel concern rather than a foreign add-on.

\textbf{Industry relevance: when the AI's output is a decision.} The
stakes sharpen outside publishing. In manufacturing, AI increasingly
prepares or makes operational decisions, releasing a weld,
classifying a part, adjusting a process window, and the worked
example of Section~\ref{sec:skill} is one step from that setting. The
same graph then answers what certification, liability, and the EU AI
Act ask of high-risk systems, whose events must be automatically
logged over the system's lifetime and which must remain under
effective human oversight~\cite{eu-ai-act2024}: which agent decided,
on which evidence, under which model version, and who signed off.
Aviation's certification guidance already anticipates exactly this
setting: EASA's AI concept paper covers systems that automatically
take decisions under human oversight and demands traceability of
data from origin to final operation through the whole pipeline,
naming record-keeping of AI-related data among the required
methodologies~\cite{easa-ai-concept2024}.
\ifextended
Space operations run the same trajectory: ESA's mission-operations
AI roadmap spans health monitoring, decision recommendation, and
planning, and names data governance and assurance as one of its five
capability pillars~\cite{esa-a2i-roadmap}, so AI-prepared operational
decisions there raise the identical who-decided-on-what-evidence
question. The framing is also recognizable to established
engineering process models: the validator is a quality gate in the
V-model sense, and the claim-to-source binding is requirements
traceability~\cite{traceability1994} applied to scholarly
assertions, so organizations operating certified quality-management
processes could adopt the graph as an evidence layer without
changing their process language.
\fi
The human-only rungs map
directly onto release decisions that must stay with people. The shepard
platform does not yet support aiprov; the integration planned in
Section~\ref{sec:conclusion} would carry the record into the same
backbone that already holds the production data.

\ifextended
\textbf{Ecosystem trajectories.} Two further consequences follow if
records like this one become common, and both are hypotheses about
use, not measured results. \emph{Training signal:} models trained on
recursively generated text degrade~\cite{model-collapse2024}; a
provenance-graphed corpus is the countervailing asset, since
attribution and verification states let curators filter training
pools by pedigree rather than fluency, and claims carrying recorded
human confirmation are precisely the scarce, high-quality labels
that data curation for scientific language models lacks today.
\emph{Contradiction detection and cascading retractions:} retracted
articles keep collecting citations long after
retraction~\cite{retracted-citations2022}; a record that binds
claims to sources makes the cascade computable, the backward closure
of Section~\ref{sec:skill}, so a retraction stops being an event at
one node and becomes a recomputation over its recorded descendants,
each affected claim flagged with \texttt{aiprov:contradicts} rather
than silently inherited.

\textbf{Philosophical footing.} The method operationalizes demands
much older than its tooling. Plato's \emph{Theaetetus} ties the
question of knowledge to justification, examining whether true
judgement counts as knowledge only with an account, a \emph{logos},
attached~\cite{plato-theaetetus}; the ladder is that demand made
structural, since every rung names who gave the account and on what
evidence, replacing trust in authority with traceable justification.
Bacon's \emph{Novum Organum} replaces the scholar's intuition with
systematic, documented procedure~\cite{bacon1620}; an aiprov record
is Baconian in exactly that sense, the logged activity rather than
the author's standing carries the epistemic weight. Kant's critical
turn holds that cognition is structured by categories the
understanding brings to experience~\cite{kant-krv}; the analogy, and
it is offered as analogy, is that the ladder, the invariants, and
claims as first-class resources are the stipulated conditions under
which the record can speak of \emph{verified} knowledge at all,
without them the graph would hold only unordered assertions. Kant's
distinction between things as they appear and as they are in
themselves also names a modesty the method practices: the record
never claims the source as such, what enters it is phenomenal
access, content-hashed vendored bytes plus the verification history,
declared as exactly that. And the universalizability test of his
practical philosophy~\cite{kant-groundwork} reads as the method's
documentation rule: record each claim so that anyone, not just its
author, could re-run the verification, which is why registry
fetches, vendoring, and generated disclosures exist. Popper makes
refutability the mark of the scientific~\cite{popper1959}; the graph
makes refutability operational rather than rhetorical, since a claim
names its sources, a reader can walk to them, and a refutation
enters the record as \texttt{aiprov:contradicts} instead of
disappearing into correspondence.

There is also a limit, and Wittgenstein names it. The
\emph{Tractatus} closes with the demand to stay silent about what
cannot be said~\cite{wittgenstein-tractatus}; read against this
method, our reading, not a claim about Wittgenstein's intent, the
last verification step is exactly such a point. What
\texttt{human-read} records is \emph{that} a named person vouched,
on a date, over identified evidence; whether that person read
carefully, understood correctly, or judged well is not itself
machine-checkable, and no additional rung could make it so, since
the same question would reattach to the new rung. The human-only
rungs are therefore not a claim that people are more reliable than
machines. They mark the boundary where structural verification ends
and attributable judgement begins, and the graph's honesty is that
it documents the boundary instead of paving over it: a recorded
refusal to promote is the method's form of silence, and the
omit-don't-estimate rule is the same discipline applied to numbers,
no invented precision where there is none. None of this is claimed
as philosophical novelty. The point is the direction of fit: the
infrastructure gives these old demands a machine-checkable form,
and marks where checkability ends, so the contribution reads not
only as a new method but as a technical answer to questions science
studies has always owned.
\fi

\textbf{Originality of this work, on both sides of the division.}
Section~\ref{sec:dilution} demands that novelty claims be bounded and
checkable; the demand applies to this paper first. So the claim is
made by the method's own mechanism: a logged novelty sweep refreshed
the search horizon, its nearest find is named and cited, and the
recorded claim is deliberately narrow. Per-activity AI provenance on
a PROV substrate is \emph{not} new, PROV-AGENT does
that~\cite{prov-agent2025}; what no found work combines is human-only
verification rungs, executable-skill packaging, and self-application,
and that combination, within the recorded horizon, is this paper's
originality claim, refutable by widening the sweep. The effort behind
it divides tellingly. The fast side, the AI's \MActiveH{} active hours,
supplied execution and recombination, and recombination is exactly
what the empirical evidence says models supply: human-written texts
contribute more new ideas to the collective pool than model-written
ones~\cite{llm-homogenization2024}. The slow side is where the
originality entered: years of experiments, specification
co-authoring, committee work, and conversations, the lineage of
Section~\ref{sec:background}, compressed into \MDirWords{} words
of direction.
Neither side alone would have produced the work: without the slow
ideas the fast hours had nothing original to execute; without the
fast execution the slow ideas would still be scattered across
repositories and conversations. The record keeps both sides visible,
which is precisely what an originality dispute would need.

\textbf{Copyright when the machine wrote most of it.} The repository's code
is Apache-2.0-licensed, but a license can only grant what
copyright first protects, and most of the repository's code and prose
was AI-generated under human direction. The U.S.\ Copyright Office
holds that copyright ``does not extend to purely AI-generated
material, or material where there is insufficient human control over
the expressive elements'', that prompts alone do not provide that
control, and that protection attaches to human-authored expression
and to ``the creative selection, coordination, or arrangement'' or
modification of outputs~\cite{usco-ai-copyrightability2025}. German
law agrees from its definition of the work: \S~2(2) UrhG protects
only \emph{pers\"onliche geistige Sch\"opfungen}~\cite{urhg-para2},
so current LLM output as such is not copyrightable in Germany. Read
against this paper's own record, the conclusion is uncomfortable and
clarifying at once: line by line, much of the artefact may be
uncopyrightable, and the license's effective scope narrows to the
human contributions and the curated whole. What the method adds is
the evidence a legal answer needs: which expression a human
conceived, selected, arranged, or modified, and which the machine
produced under how much control, is exactly what the per-activity
graph records (Table~\ref{tab:division}) and what a blanket
``AI-assisted'' disclaimer erases. Rights analysis of AI-era works
founders on evidence before doctrine; provenance-first authoring is
how the evidence survives.

\textbf{Nostalgia for the paper, or the wrong artefact instrumented?}
The sharpest objection to this work is not about any rung but about
its target: perhaps F(AI)\textsuperscript{2}R merely patches
provenance onto a form we are nostalgic for, while AI-driven science
becomes something structurally different: living claim graphs,
continuously re-executed analyses, agents exchanging machine-readable
evidence, narrative generated on demand. If so, instrumenting PDF
production is polishing the wrong artefact. But the method is less
paper-bound than its demonstration: the invariants and ladder attach
to claims, sources, activities, and agents, never to a manuscript,
and here the graph is the primary record, the PDF one derived view
CI rebuilds from it. Read that
way, F(AI)\textsuperscript{2}R is a migration path, moving a
publication's substance into the claim-level form a post-paper
science would need while still emitting the legacy format. If the
paper form dissolves, judge this method by whether its graphs remain
useful without the PDFs.

\textbf{The force multiplier, both ways, and the optimistic reading.}
One operator ran method, paper, and integration fork in parallel
during this experiment; unaccounted, the same multiplication is the
dilution engine of Section~\ref{sec:dilution}. The differentiator is
not speed but auditability: fast work that carries its record is
distinguishable from fast noise. Fast paper, slow ideas is the honest
reading of this project, and only the graph makes that reading
checkable. There is also a hopeful version of the same observation.
What the multiplication buys is not mainly volume: AI absorbing the
tedium (registry lookups, hash binding, conformance checks, build
plumbing) can free human attention for the two things machines do not
supply, connecting ideas and connecting people. This experiment's own
activity log shows that division of labour. No guarantee, only a
chance; a provenance graph is how one would check, later, whether the
chance was taken. For this experiment the leverage measurement of
Section~\ref{sec:meta} is that check, taken early: \MDirWords{}
typed words steering \MActiveH{} active hours, nearly all of them
spent on scope, semantics, lineage, or people.

\section{Conclusion}\label{sec:conclusion}

We generalized the F(AI)\textsuperscript{2}R method into aiprov, a
domain-agnostic PROV-O extension for AI-in-the-loop work; we packaged
the method as an executable skill an AI agent operates itself; and we
applied it to its own production, mishaps, mid-experiment redesigns,
and all. The stance of the original experiment carries over and
extends: the repository is the paper, is the process, and now also is
the method, versioned, validated, and shipped alongside what it
produced.

Three roots converge here: AI-assisted research practice, which
contributed transcript-as-artifact; engineering provenance, which
insists on traceability; and industrial semantic modeling, which
taught that where exchange matters, semantics must be formal. Where
semantics are important, provenance is the semantics of work.

The path forward is infrastructure. shepard~\cite{shepard}, a jointly
developed DLR system for heterogeneous product and research data,
already serves as the data backbone of engineering deployments such as
the MEMAS additive-manufacturing pipeline~\cite{memas2024}; an
experimental fork~\cite{shepard-fork} is the staging ground for
integrating aiprov into it as planned work. Where shepard instances
capture what was produced and measured, aiprov adds who, human or AI,
shaped it and how it was verified. Further work includes multi-agent
attribution, energy accounting, registry integration beyond ORCID and
DOI, vendoring sources at the access gate by default, a combined
profile with uncertainty-aware provenance~\cite{uncertainty-prov2026},
and registering the vocabulary namespace as a permanent identifier,
deliberately deferred so that community feedback shapes the terms
before the identifier freezes.
Two further items come straight from reviewing this work against its
own standard. The \texttt{ai-confirmed} judgements here were granted
by the agent that wrote the citing sentences; measuring that
self-confirmation bias, by having an independent model or sampled
humans re-verify the same sources, is the next audit the
method owes itself. And the meta-experiment is a single self-applied
case; the portability claims of Section~\ref{sec:discussion} earn
their keep only when a second operator applies the skill in a second
domain and the records are compared.

\ifextended
The integration path scales beyond a single repository. An aiprov
record is a FAIR digital object in waiting: typed, persistently
identified, machine-actionable~\cite{fairdo2020}. Binding it to
kernel information profiles~\cite{hmc-kip2022} would let
infrastructure such as the Helmholtz Knowledge
Graph~\cite{unhide-helmholtz-kg} harvest AI-work provenance exactly
as it harvests dataset metadata today, and the shepard integration
named above is the concrete first edge of that graph. The vision is
ecosystem-level rather than paper-local: provenance as a backbone on
which AI-assisted science can be audited, disputed, and credited
across institutions, with the skill as the unit that makes the
practice installable and the reviewer-library, backward-closure, and
dispute designs of Section~\ref{sec:skill} as the next increments.

\fi
One thing must be said with urgency, and plainly. Nothing in this
paper settles what science with AI in the loop should become. That
understanding is being negotiated now, by editors rewriting authorship
policies, regulators phasing in transparency duties, funders and
communities arguing over credit, originality, cost, and trust, and it
will be settled socially, not technically. This initiative is not the
solution; it is technical support for whichever understanding society
arrives at, a way to keep the record straight while the norms are
still moving. But the discussion itself cannot wait, because
infrastructure that ships first tends to settle norms by default, and
the most expensive mistake would be to let tools decide what the
community has not yet debated. We ask for that debate, and we offer
this repository's record, its graph, its transcript, its incidents,
its bill, as material evidence to hold it over.

\section*{Acknowledgment and AI Transparency Statement}
The ideas in this paper were sharpened through earlier
experiments~\cite{obscurity-is-dead, fair2r-repo}; through discussions
in the context of the Helmholtz Metadata Collaboration
(HMC)~\cite{hmc}: with Witold Arndt in the HMC hub for
aeronautics, space and transport, and through the broad view of AI in
research afforded by chairing the HMC Conference 2025 programme
committee~\cite{hmc-conference2025}; through exchanges with Frank
Dressel on provenance for engineering
systems~\cite{uncertainty-prov2026}; through deep discussions with
Sirko Schindler and Carsten Hoyer-Klick on ontologies; through the
research-data-management evangelism of Christian Langenbach at DLR;
and through the inspiration of Carina Haupt's work on provenance
use cases~\cite{provenance-overview2022} and of the MEMAS project's
integrated data management~\cite{memas2024}.

This paper and its repository were produced with an AI agent operating
the ai-provenance skill; every activity, generated artefact, claim, and
source is recorded in the version-controlled provenance graph,
including the production of this statement. The following disclosure is
generated from that graph by \texttt{provlog.py disclosure} and is made
in view of the transparency rules of the EU Artificial Intelligence
Act~\cite{eu-ai-act2024}:

\begingroup\itshape
Parts of this work were generated or edited with the assistance of artificial-intelligence systems: Claude Code (Claude Fable 5, Anthropic), associated with 206 recorded activities. In line with the transparency obligations for AI-generated content under Regulation (EU) 2024/1689 (AI Act), this assistance is disclosed here and is additionally marked in machine-readable form: the version-controlled provenance graph (provenance.ttl, W3C PROV-O) records the 244 activities behind this work, the 61 generated artefacts with content hashes, and the 8 recorded claims with their verification states. Human oversight is structural: verification rungs above ai-confirmed are reserved to human agents, and the conformance validator rejects AI-granted promotions.

\endgroup

\emph{Availability:} skill bundle, provenance graph, transcripts,
and paper sources are public at
\url{https://github.com/noheton/f-ai2-r}; the skill loads in any
Agent Skills client, and conformance-gated releases carry bundle,
PDF, graph, and checksums. The data underlying the demonstration is
archived on Zenodo (DOI 10.5281/zenodo.21667684).

\bibliographystyle{IEEEtran}
\bibliography{references}

% Generated by IEEEtran.bst, version: 1.14 (2015/08/26)
\begin{thebibliography}{10}
\providecommand{\url}[1]{#1}
\csname url@samestyle\endcsname
\providecommand{\newblock}{\relax}
\providecommand{\bibinfo}[2]{#2}
\providecommand{\BIBentrySTDinterwordspacing}{\spaceskip=0pt\relax}
\providecommand{\BIBentryALTinterwordstretchfactor}{4}
\providecommand{\BIBentryALTinterwordspacing}{\spaceskip=\fontdimen2\font plus
\BIBentryALTinterwordstretchfactor\fontdimen3\font minus
  \fontdimen4\font\relax}
\providecommand{\BIBforeignlanguage}[2]{{%
\expandafter\ifx\csname l@#1\endcsname\relax
\typeout{** WARNING: IEEEtran.bst: No hyphenation pattern has been}%
\typeout{** loaded for the language `#1'. Using the pattern for}%
\typeout{** the default language instead.}%
\else
\language=\csname l@#1\endcsname
\fi
#2}}
\providecommand{\BIBdecl}{\relax}
\BIBdecl

\bibitem{fair2016}
\BIBentryALTinterwordspacing
M.~D. Wilkinson, M.~Dumontier \emph{et~al.}, ``The {FAIR} guiding principles
  for scientific data management and stewardship,'' \emph{Scientific Data},
  vol.~3, no.~1, Mar. 2016. [Online]. Available:
  \url{http://dx.doi.org/10.1038/sdata.2016.18}
\BIBentrySTDinterwordspacing

\bibitem{chatgpt-wrote-it2023}
\BIBentryALTinterwordspacing
Y.~K. Dwivedi \emph{et~al.}, ``Opinion paper: “so what if {ChatGPT} wrote
  it?” multidisciplinary perspectives on opportunities, challenges and
  implications of generative conversational {AI} for research, practice and
  policy,'' \emph{International Journal of Information Management}, vol.~71, p.
  102642, Aug. 2023. [Online]. Available:
  \url{http://dx.doi.org/10.1016/j.ijinfomgt.2023.102642}
\BIBentrySTDinterwordspacing

\bibitem{chatgpt-priorities2023}
\BIBentryALTinterwordspacing
E.~A.~M. van Dis, J.~Bollen \emph{et~al.}, ``{ChatGPT}: five priorities for
  research,'' \emph{Nature}, vol. 614, no. 7947, pp. 224--226, Feb. 2023.
  [Online]. Available: \url{http://dx.doi.org/10.1038/d41586-023-00288-7}
\BIBentrySTDinterwordspacing

\bibitem{modelcards2019}
\BIBentryALTinterwordspacing
M.~Mitchell, S.~Wu \emph{et~al.}, ``Model cards for model reporting,'' in
  \emph{Proceedings of the Conference on Fairness, Accountability, and
  Transparency}, ser. FAT* ’19.\hskip 1em plus 0.5em minus 0.4em\relax ACM,
  Jan. 2019, pp. 220--229. [Online]. Available:
  \url{http://dx.doi.org/10.1145/3287560.3287596}
\BIBentrySTDinterwordspacing

\bibitem{perverse-incentives2017}
\BIBentryALTinterwordspacing
M.~A. Edwards and S.~Roy, ``Academic research in the 21st century: Maintaining
  scientific integrity in a climate of perverse incentives and
  hypercompetition,'' \emph{Environmental Engineering Science}, vol.~34, no.~1,
  pp. 51--61, Jan. 2017. [Online]. Available:
  \url{http://dx.doi.org/10.1089/ees.2016.0223}
\BIBentrySTDinterwordspacing

\bibitem{publishing-strain2024}
\BIBentryALTinterwordspacing
M.~A. Hanson, P.~G. Barreiro, P.~Crosetto, and D.~Brockington, ``The strain on
  scientific publishing,'' \emph{Quantitative Science Studies}, vol.~5, no.~4,
  pp. 823--843, 2024. [Online]. Available:
  \url{http://dx.doi.org/10.1162/qss_a_00327}
\BIBentrySTDinterwordspacing

\bibitem{obscurity-is-dead}
\BIBentryALTinterwordspacing
F.~Krebs, ``Obscurity is dead --- proprietary by design, open by {AI},''
  Research repository with evidentiary artifacts, n.d. [Online]. Available:
  \url{https://github.com/noheton/Obscurity-Is-Dead}
\BIBentrySTDinterwordspacing

\bibitem{fair2r-repo}
\BIBentryALTinterwordspacing
------, ``{F(AI)\textsuperscript{2}R}: {FAIR} research with {AI} in the loop,
  twice,'' Research repository: manuscript, provenance graph, transcripts, n.d.
  [Online]. Available: \url{https://github.com/noheton/f-ai-r}
\BIBentrySTDinterwordspacing

\bibitem{shepard-fork}
\BIBentryALTinterwordspacing
------, ``Experimental fork of shepard,'' Development/research workspace with
  experimental v2 endpoints, n.d. [Online]. Available:
  \url{https://github.com/noheton/shepard}
\BIBentrySTDinterwordspacing

\bibitem{prov-o2013}
\BIBentryALTinterwordspacing
T.~Lebo, S.~Sahoo, and D.~McGuinness, ``{PROV-O}: The {PROV} ontology,'' W3C
  Recommendation, Apr. 2013, accessed 2026-07-24. [Online]. Available:
  \url{https://www.w3.org/TR/prov-o/}
\BIBentrySTDinterwordspacing

\bibitem{opm2010}
\BIBentryALTinterwordspacing
L.~Moreau, B.~Clifford \emph{et~al.}, ``The open provenance model core
  specification (v1.1),'' \emph{Future Generation Computer Systems}, vol.~27,
  no.~6, pp. 743--756, Jun. 2011. [Online]. Available:
  \url{http://dx.doi.org/10.1016/j.future.2010.07.005}
\BIBentrySTDinterwordspacing

\bibitem{pav2013}
\BIBentryALTinterwordspacing
P.~Ciccarese, S.~Soiland-Reyes, K.~Belhajjame, A.~J. Gray, C.~Goble, and
  T.~Clark, ``{PAV} ontology: provenance, authoring and versioning,''
  \emph{Journal of Biomedical Semantics}, vol.~4, no.~1, p.~37, 2013. [Online].
  Available: \url{http://dx.doi.org/10.1186/2041-1480-4-37}
\BIBentrySTDinterwordspacing

\bibitem{provenance-overview2022}
\BIBentryALTinterwordspacing
C.~Haupt, ``An overview of provenance and its use cases,'' RDA Deutschland
  Tagung 2022; DLR elib 185438, 2022. [Online]. Available:
  \url{https://elib.dlr.de/185438/}
\BIBentrySTDinterwordspacing

\bibitem{credit2019}
\BIBentryALTinterwordspacing
L.~Allen, A.~O’Connell, and V.~Kiermer, ``How can we ensure visibility and
  diversity in research contributions? how the contributor role taxonomy
  ({CRediT}) is helping the shift from authorship to contributorship,''
  \emph{Learned Publishing}, vol.~32, no.~1, pp. 71--74, Jan. 2019. [Online].
  Available: \url{http://dx.doi.org/10.1002/leap.1210}
\BIBentrySTDinterwordspacing

\bibitem{credit-niso2022}
\BIBentryALTinterwordspacing
{NISO}, ``{ANSI/NISO Z39.104-2022, CRediT, Contributor Roles Taxonomy},'' 2022.
  [Online]. Available: \url{http://dx.doi.org/10.3789/ansi.niso.z39.104-2022}
\BIBentrySTDinterwordspacing

\bibitem{datasheets2021}
\BIBentryALTinterwordspacing
T.~Gebru, J.~Morgenstern \emph{et~al.}, ``Datasheets for datasets,''
  \emph{Communications of the ACM}, vol.~64, no.~12, pp. 86--92, Nov. 2021.
  [Online]. Available: \url{http://dx.doi.org/10.1145/3458723}
\BIBentrySTDinterwordspacing

\bibitem{rocrate2022}
\BIBentryALTinterwordspacing
S.~Soiland-Reyes, P.~Sefton \emph{et~al.}, ``Packaging research artefacts with
  {RO-Crate},'' \emph{Data Science}, vol.~5, no.~2, pp. 97--138, Jan. 2022.
  [Online]. Available: \url{http://dx.doi.org/10.3233/ds-210053}
\BIBentrySTDinterwordspacing

\bibitem{prov-agent2025}
\BIBentryALTinterwordspacing
R.~Souza, A.~Gueroudji, S.~DeWitt, D.~Rosendo, T.~Ghosal, R.~Ross,
  P.~Balaprakash, and R.~F. Da~Silva, ``{PROV-AGENT}: Unified provenance for
  tracking {AI} agent interactions in agentic workflows,'' in \emph{2025 IEEE
  International Conference on eScience (eScience)}.\hskip 1em plus 0.5em minus
  0.4em\relax IEEE, Sep. 2025, pp. 467--473. [Online]. Available:
  \url{http://dx.doi.org/10.1109/escience65000.2025.00093}
\BIBentrySTDinterwordspacing

\bibitem{fairdo2020}
\BIBentryALTinterwordspacing
K.~De~Smedt, D.~Koureas, and P.~Wittenburg, ``{FAIR} digital objects for
  science: From data pieces to actionable knowledge units,''
  \emph{Publications}, vol.~8, no.~2, p.~21, Apr. 2020. [Online]. Available:
  \url{http://dx.doi.org/10.3390/publications8020021}
\BIBentrySTDinterwordspacing

\bibitem{hmc-kip2022}
\BIBentryALTinterwordspacing
{Helmholtz Metadata Collaboration} and C.~Curdt, ``Helmholtz metadata
  collaboration, helmholtz kernel information profile,'' HMC Office, GEOMAR
  Helmholtz Centre for Ocean Research Kiel, Tech. Rep., 2022. [Online].
  Available: \url{http://dx.doi.org/10.3289/hmc_publ_03}
\BIBentrySTDinterwordspacing

\bibitem{unhide-helmholtz-kg}
\BIBentryALTinterwordspacing
J.~Bröder, G.~Preuß, F.~D’Mello, S.~Fathalla, V.~Hofmann, and S.~Sandfeld,
  \emph{The Helmholtz Knowledge Graph: Driving the Transition Towards a FAIR
  Data Ecosystem in the Helmholtz Association}.\hskip 1em plus 0.5em minus
  0.4em\relax Springer Nature Switzerland, 2025, pp. 183--187. [Online].
  Available: \url{http://dx.doi.org/10.1007/978-3-031-78952-6_23}
\BIBentrySTDinterwordspacing

\bibitem{attention2017}
\BIBentryALTinterwordspacing
A.~Vaswani, N.~Shazeer, N.~Parmar, J.~Uszkoreit, L.~Jones, A.~N. Gomez,
  L.~Kaiser, and I.~Polosukhin, ``Attention is all you need,'' arXiv preprint
  arXiv:1706.03762, 2017. [Online]. Available:
  \url{https://arxiv.org/abs/1706.03762}
\BIBentrySTDinterwordspacing

\bibitem{ai-hallucination2023}
\BIBentryALTinterwordspacing
H.~Alkaissi and S.~I. McFarlane, ``Artificial hallucinations in {ChatGPT}:
  Implications in scientific writing,'' \emph{Cureus}, Feb. 2023. [Online].
  Available: \url{http://dx.doi.org/10.7759/cureus.35179}
\BIBentrySTDinterwordspacing

\bibitem{nanopub2010}
\BIBentryALTinterwordspacing
P.~Groth, A.~Gibson, and J.~Velterop, ``The anatomy of a nanopublication,''
  \emph{Information Services and Use}, vol.~30, no. 1-2, pp. 51--56, Feb. 2010.
  [Online]. Available: \url{http://dx.doi.org/10.3233/ISU-2010-0613}
\BIBentrySTDinterwordspacing

\bibitem{micropub2014}
\BIBentryALTinterwordspacing
T.~Clark, P.~N. Ciccarese, and C.~A. Goble, ``Micropublications: a semantic
  model for claims, evidence, arguments and annotations in biomedical
  communications,'' \emph{Journal of Biomedical Semantics}, vol.~5, no.~1,
  p.~28, 2014. [Online]. Available:
  \url{http://dx.doi.org/10.1186/2041-1480-5-28}
\BIBentrySTDinterwordspacing

\bibitem{c2pa-spec}
\BIBentryALTinterwordspacing
{Coalition for Content Provenance and Authenticity}, ``Content credentials:
  {C2PA} technical specification,'' Specification v2.4, 2025, accessed
  2026-07-28. [Online]. Available: \url{https://c2pa.org/specifications/}
\BIBentrySTDinterwordspacing

\bibitem{in-toto2019}
\BIBentryALTinterwordspacing
S.~Torres-Arias, H.~Afzali, T.~K. Kuppusamy, R.~Curtmola, and J.~Cappos,
  ``in-toto: Providing farm-to-table guarantees for bits and bytes,'' in
  \emph{28th USENIX Security Symposium}, 2019. [Online]. Available:
  \url{https://www.usenix.org/conference/usenixsecurity19/presentation/torres-arias}
\BIBentrySTDinterwordspacing

\bibitem{manubot2019}
\BIBentryALTinterwordspacing
D.~S. Himmelstein, V.~Rubinetti \emph{et~al.}, ``Open collaborative writing
  with {Manubot},'' \emph{PLOS Computational Biology}, vol.~15, no.~6, p.
  e1007128, Jun. 2019. [Online]. Available:
  \url{http://dx.doi.org/10.1371/journal.pcbi.1007128}
\BIBentrySTDinterwordspacing

\bibitem{uncertainty-prov2026}
\BIBentryALTinterwordspacing
D.~Valente, A.~Sch\"afer, E.~Tasdemir, R.~Hoppe, O.~Bertram, and F.~Dressel,
  ``An uncertainty-aware provenance framework for enhanced traceability in
  engineering systems,'' DLR elib 224241; submitted to IEEE Aerospace and
  Electronic Systems Magazine, 2026. [Online]. Available:
  \url{https://elib.dlr.de/224241/}
\BIBentrySTDinterwordspacing

\bibitem{better-architecture2025}
\BIBentryALTinterwordspacing
S.~Druskat, N.~U. Eisty, R.~Chisholm, N.~P. Chue~Hong, R.~C. Cocking, M.~B.
  Cohen, M.~Felderer, L.~Grunske, S.~A. Harris, W.~Hasselbring, T.~Krause,
  J.~Linxweiler, and C.~C. Venters, ``Better architecture, better software,
  better research,'' \emph{Computing in Science \& Engineering}, vol.~27,
  no.~2, pp. 45--57, Apr. 2025. [Online]. Available:
  \url{http://dx.doi.org/10.1109/MCSE.2025.3573887}
\BIBentrySTDinterwordspacing

\bibitem{aas-part1-v30rc02}
\BIBentryALTinterwordspacing
{Plattform Industrie 4.0}, ``Details of the asset administration shell. {Part}
  1 --- the exchange of information between partners in the value chain of
  {Industrie} 4.0 ({Version} 3.0{RC}02),'' Federal Ministry for Economic
  Affairs and Climate Action (BMWK), specification, 2022. [Online]. Available:
  \url{https://www.plattform-i40.de/IP/Redaktion/EN/Downloads/Publikation/Details_of_the_Asset_Administration_Shell_Part1_V3.html}
\BIBentrySTDinterwordspacing

\bibitem{aas-part1-idta}
\BIBentryALTinterwordspacing
{Industrial Digital Twin Association (IDTA)}, ``Specification of the asset
  administration shell. {Part} 1: Metamodel,'' IDTA-01001, n.d. [Online].
  Available: \url{https://doi.org/10.62628/idta.01001-3-1-1}
\BIBentrySTDinterwordspacing

\bibitem{hmc}
\BIBentryALTinterwordspacing
{Helmholtz Association}, ``Helmholtz metadata collaboration ({HMC}),'' n.d.
  [Online]. Available: \url{https://helmholtz-metadaten.de/}
\BIBentrySTDinterwordspacing

\bibitem{hmc-conference2025}
\BIBentryALTinterwordspacing
{Helmholtz Metadata Collaboration}, ``{HMC} conference 2025 --- book of
  abstracts,'' 2025. [Online]. Available:
  \url{https://helmholtz-metadaten.de/storage/2298/HMC-Conference_2025_Book-of-Abstracts.pdf}
\BIBentrySTDinterwordspacing

\bibitem{paper-mills2021}
\BIBentryALTinterwordspacing
S.~Heck, F.~Bianchini \emph{et~al.}, ``Fake data, paper mills, and their
  authors: The {International Journal of Cancer} reacts to this threat to
  scientific integrity,'' \emph{International Journal of Cancer}, vol. 149,
  no.~3, pp. 492--493, Apr. 2021. [Online]. Available:
  \url{http://dx.doi.org/10.1002/ijc.33604}
\BIBentrySTDinterwordspacing

\bibitem{gpt-fabricated2024}
\BIBentryALTinterwordspacing
J.~Haider, K.~R. Söderström, B.~Ekström, and M.~Rödl, ``{GPT}-fabricated
  scientific papers on {Google Scholar}: Key features, spread, and implications
  for preempting evidence manipulation,'' \emph{Harvard Kennedy School
  Misinformation Review}, Sep. 2024. [Online]. Available:
  \url{http://dx.doi.org/10.37016/mr-2020-156}
\BIBentrySTDinterwordspacing

\bibitem{fabricated-citations2023}
\BIBentryALTinterwordspacing
W.~H. Walters and E.~I. Wilder, ``Fabrication and errors in the bibliographic
  citations generated by {ChatGPT},'' \emph{Scientific Reports}, vol.~13,
  no.~1, Sep. 2023. [Online]. Available:
  \url{http://dx.doi.org/10.1038/s41598-023-41032-5}
\BIBentrySTDinterwordspacing

\bibitem{model-collapse2024}
\BIBentryALTinterwordspacing
I.~Shumailov, Z.~Shumaylov \emph{et~al.}, ``{AI} models collapse when trained
  on recursively generated data,'' \emph{Nature}, vol. 631, no. 8022, pp.
  755--759, Jul. 2024. [Online]. Available:
  \url{http://dx.doi.org/10.1038/s41586-024-07566-y}
\BIBentrySTDinterwordspacing

\bibitem{eu-ai-act2024}
\BIBentryALTinterwordspacing
{European Parliament and Council of the European Union}, ``Regulation ({EU})
  2024/1689 of the {European} {Parliament} and of the {Council} laying down
  harmonised rules on artificial intelligence ({Artificial} {Intelligence}
  {Act}),'' Official Journal of the European Union, L series, 2024, accessed
  2026-07-24. [Online]. Available:
  \url{https://eur-lex.europa.eu/eli/reg/2024/1689/oj}
\BIBentrySTDinterwordspacing

\bibitem{ultrasonic-welding2025}
\BIBentryALTinterwordspacing
M.~Janek, D.~Görick, L.~Larsen, S.~Jarka, and M.~Kupke, ``Investigation of
  power and amplitude control in continuous ultrasonic welding of
  unidirectional {CFRPs}: A comparative study,'' \emph{Composites Part A:
  Applied Science and Manufacturing}, vol. 199, p. 109194, Dec. 2025. [Online].
  Available: \url{http://dx.doi.org/10.1016/j.compositesa.2025.109194}
\BIBentrySTDinterwordspacing

\bibitem{anthropic-pricing}
\BIBentryALTinterwordspacing
{Anthropic}, ``Claude {API} pricing,'' Developer documentation, 2026, accessed
  2026-07-28. [Online]. Available:
  \url{https://platform.claude.com/docs/en/about-claude/pricing}
\BIBentrySTDinterwordspacing

\bibitem{doctors-vs-algorithms2026}
\BIBentryALTinterwordspacing
A.~Vinas, F.~Blanco, and H.~Matute, ``Doctors vs. algorithms: Physicians, too,
  struggle to learn from evidence that contradicts {AI} suggestions,''
  \emph{PLOS Digital Health}, vol.~5, no.~7, p. e0001490, Jul. 2026. [Online].
  Available: \url{http://dx.doi.org/10.1371/journal.pdig.0001490}
\BIBentrySTDinterwordspacing

\bibitem{eu-ai-icons2026}
{European Commission}, ``{EU} icons for labelling {AI}-generated content,''
  \url{https://digital-strategy.ec.europa.eu/en/policies/eu-icons-labelling-ai-generated-content},
  Jul. 2026, freely usable icons supporting Article 50(4) {AI} {Act}
  disclosure; part of the Code of Practice on marking and labelling of
  {AI}-generated content.

\bibitem{faq-ai-gwp2025}
\BIBentryALTinterwordspacing
K.~Frisch, ``{FAQ} {K}\"unstliche {I}ntelligenz und gute wissenschaftliche
  {P}raxis, {Version} 2,'' Oct. 2025, fAQ from the German research-integrity
  ombuds context; includes the English version 2.1, {FAQ} {AI} and Research
  Integrity. [Online]. Available: \url{https://doi.org/10.5281/zenodo.17349995}
\BIBentrySTDinterwordspacing

\bibitem{oecd-ai-principles}
\BIBentryALTinterwordspacing
{OECD}, ``Recommendation of the {Council} on {Artificial} {Intelligence}
  ({OECD}/{LEGAL}-0449),'' Adopted May 2019, amended May 2024, 2024, accessed
  2026-07-27. [Online]. Available: \url{https://oecd.ai/en/ai-principles}
\BIBentrySTDinterwordspacing

\bibitem{oecd-aim}
\BIBentryALTinterwordspacing
{OECD.AI Policy Observatory}, ``{AI} incidents and hazards monitor ({AIM}),''
  2024, accessed 2026-07-27. [Online]. Available:
  \url{https://oecd.ai/en/incidents-methodology}
\BIBentrySTDinterwordspacing

\bibitem{agent-skills-spec}
{Anthropic}, ``Agent skills: An open format for extending {AI} agent
  capabilities,'' \url{https://agentskills.io/}, 2025, specification and
  adopter list; originally developed by Anthropic and released as an open
  standard.

\bibitem{easa-ai-concept2024}
{European Union Aviation Safety Agency}, ``{EASA} artificial intelligence
  concept paper issue 2: Guidance for level 1 \& 2 machine-learning
  applications,''
  \url{https://www.easa.europa.eu/en/document-library/general-publications/easa-artificial-intelligence-concept-paper-issue-2},
  Mar. 2024.

\bibitem{esa-a2i-roadmap}
{European Space Agency, ESOC}, ``{A\textsuperscript{2}I} roadmap at {ESA}'s
  mission operations,''
  \url{https://esoc.esa.int/a2i-roadmap-esas-missions-operations}, 2026,
  artificial Intelligence for Automation roadmap: five priority domains and
  fourteen use cases in mission operations; developed at ESOC with an industry
  consortium.

\bibitem{traceability1994}
\BIBentryALTinterwordspacing
O.~Gotel and C.~Finkelstein, ``An analysis of the requirements traceability
  problem,'' in \emph{Proceedings of IEEE International Conference on
  Requirements Engineering}, ser. ICRE-94.\hskip 1em plus 0.5em minus
  0.4em\relax IEEE Comput. Soc. Press, 1994, pp. 94--101. [Online]. Available:
  \url{http://dx.doi.org/10.1109/icre.1994.292398}
\BIBentrySTDinterwordspacing

\bibitem{retracted-citations2022}
\BIBentryALTinterwordspacing
I.~Heibi and S.~Peroni, ``A quantitative and qualitative open citation analysis
  of retracted articles in the humanities,'' \emph{Quantitative Science
  Studies}, vol.~3, no.~4, pp. 953--975, 2022. [Online]. Available:
  \url{http://dx.doi.org/10.1162/qss_a_00222}
\BIBentrySTDinterwordspacing

\bibitem{plato-theaetetus}
\BIBentryALTinterwordspacing
Plato, \emph{Theaetetus}, ser. Clarendon Plato Series.\hskip 1em plus 0.5em
  minus 0.4em\relax Oxford University Press, 1973, transl. and ed. McDowell.
  [Online]. Available:
  \url{http://dx.doi.org/10.1093/actrade/9780198720836.book.1}
\BIBentrySTDinterwordspacing

\bibitem{bacon1620}
\BIBentryALTinterwordspacing
F.~Bacon, \emph{Novum Organum}.\hskip 1em plus 0.5em minus 0.4em\relax Oxford
  University Press, 1620, pp. 48--586, ed. Rees and Wakely. [Online].
  Available: \url{http://dx.doi.org/10.1093/oseo/instance.00007242}
\BIBentrySTDinterwordspacing

\bibitem{kant-krv}
\BIBentryALTinterwordspacing
I.~Kant, \emph{Critique of Pure Reason}.\hskip 1em plus 0.5em minus 0.4em\relax
  Cambridge University Press, 1998, first published 1781/1787; ed. and transl.
  Guyer and Wood. [Online]. Available:
  \url{http://dx.doi.org/10.1017/CBO9780511804649}
\BIBentrySTDinterwordspacing

\bibitem{kant-groundwork}
\BIBentryALTinterwordspacing
------, \emph{Groundwork of the Metaphysics of Morals}.\hskip 1em plus 0.5em
  minus 0.4em\relax Cambridge University Press, 1998, first published 1785;
  transl. Gregor, introd. Korsgaard. [Online]. Available:
  \url{http://dx.doi.org/10.1017/CBO9780511809590}
\BIBentrySTDinterwordspacing

\bibitem{popper1959}
\BIBentryALTinterwordspacing
K.~Popper, \emph{The Logic of Scientific Discovery}.\hskip 1em plus 0.5em minus
  0.4em\relax Routledge, 2005, first published as Logik der Forschung, 1934;
  English edition 1959. [Online]. Available:
  \url{http://dx.doi.org/10.4324/9780203994627}
\BIBentrySTDinterwordspacing

\bibitem{wittgenstein-tractatus}
\BIBentryALTinterwordspacing
L.~Wittgenstein, \emph{Tractatus Logico-Philosophicus}.\hskip 1em plus 0.5em
  minus 0.4em\relax Routledge, 2003, first published 1921. [Online]. Available:
  \url{http://dx.doi.org/10.4324/9780203010341}
\BIBentrySTDinterwordspacing

\bibitem{llm-homogenization2024}
\BIBentryALTinterwordspacing
K.~Moon, A.~Green, and K.~Kushlev, ``Homogenizing effect of a large language
  model ({LLM}) on creative diversity: An empirical comparison of human and
  {ChatGPT} writing, preprint,'' Aug. 2024. [Online]. Available:
  \url{http://dx.doi.org/10.31234/osf.io/8p9wu}
\BIBentrySTDinterwordspacing

\bibitem{usco-ai-copyrightability2025}
{U.S. Copyright Office}, ``Copyright and artificial intelligence, part 2:
  Copyrightability,'' Report of the Register of Copyrights,
  \url{https://www.copyright.gov/ai/}, Jan. 2025.

\bibitem{urhg-para2}
{Bundesrepublik Deutschland}, ``{Gesetz {\"u}ber Urheberrecht und verwandte
  Schutzrechte (UrhG), {\S}\,2 Gesch{\"u}tzte Werke},''
  \url{https://www.gesetze-im-internet.de/urhg/__2.html}, 1965.

\bibitem{shepard}
\BIBentryALTinterwordspacing
T.~Haase, R.~Gl{\"u}ck, P.~Kaufmann, and M.~Willmeroth, ``shepard: storage for
  heterogeneous product and research data,'' Zenodo, v5.1.2, DLR, Dec. 2025.
  [Online]. Available: \url{https://doi.org/10.5281/zenodo.17897485}
\BIBentrySTDinterwordspacing

\bibitem{memas2024}
\BIBentryALTinterwordspacing
N.~Unger, P.~Kamble, M.~Vinot, and R.~Gl\"uck, ``Project {MEMAS}: Integrated
  data management for additive manufacturing enabling high-fidelity modeling,''
  HMC Conference 2024, 2024. [Online]. Available:
  \url{https://elib.dlr.de/208404/}
\BIBentrySTDinterwordspacing

\end{thebibliography}

\typeout{get arXiv to do 4 passes: Label(s) may have changed. Rerun}
\end{document}